\documentclass[twocolumn]{aastex631}
\usepackage{stfloats}
\usepackage{float}
\usepackage{graphicx}
\usepackage{natbib}
\usepackage{hyperref}
\usepackage{amsmath}
\usepackage{autobreak}
\usepackage{multirow}
\usepackage{makecell}
\usepackage{color}
\usepackage{bm}
\usepackage{threeparttable}
\usepackage[figuresright]{rotating}
\usepackage[mathscr]{euscript}
\usepackage[mathlines]{lineno}
\usepackage[utf8]{inputenc}
\usepackage[T1]{fontenc}
\usepackage{amsmath}
\usepackage{makecell}
\usepackage{subfigure}
\usepackage{rotating}
\usepackage{CJKutf8}

\begin{document}

\correspondingauthor{Ning-Chen Sun}
\email{sunnc@ucas.ac.cn}

\author[0000-0002-3651-0681]{Zexi Niu}
\affiliation{School of Astronomy and Space Science, University of Chinese Academy of Sciences, Beijing 100049, People's Republic of China}
\affiliation{National Astronomical Observatories, Chinese Academy of Sciences, Beijing 100101, China}

\author{Ning-Chen Sun}
\affiliation{School of Astronomy and Space Science, University of Chinese Academy of Sciences, Beijing 100049, People's Republic of China}
\affiliation{National Astronomical Observatories, Chinese Academy of Sciences, Beijing 100101, China}
\affiliation{Institute for Frontiers in Astronomy and Astrophysics, Beijing Normal University, Beijing, 102206, People's Republic of China}

\author[0000-0002-7464-498X]{Emmanouil Zapartas}
\affiliation{Institute of Astrophysics, Foundation for Research and Technology-Hellas, GR-71110 Heraklion, Greece}
\affiliation{Physics Department, National and Kapodistrian University of Athens, 15784 Athens, Greece}

\author[0000-0003-4175-4960]{Conor L. Ransome}
\affiliation{Steward Observatory, University of Arizona, 933 N. Cherry Street, Tucson, AZ 85721, USA}

\author[0000-0003-0733-7215]{Justyn R. Maund}
\affiliation{Department of Physics, Royal Holloway, University of London, Egham, TW20 0EX, United Kingdom}

\author[0000-0002-7559-315X]{Cesar Rojas-Bravo}
\affiliation{School of Astronomy and Space Science, University of Chinese Academy of Sciences, Beijing 100049, People's Republic of China}
\affiliation{National Astronomical Observatories, Chinese Academy of Sciences, Beijing 100101, China}

\author{Jifeng Liu}
\affiliation{National Astronomical Observatories, Chinese Academy of Sciences, Beijing 100101, China}
\affiliation{School of Astronomy and Space Science, University of Chinese Academy of Sciences, Beijing 100049, People's Republic of China}
\affiliation{Institute for Frontiers in Astronomy and Astrophysics, Beijing Normal University, Beijing, 102206, People's Republic of China}
\affiliation{New Cornerstone Science Laboratory, National Astronomical Observatories, Chinese Academy of Sciences, Beijing 100012, People's Republic of China}

\title{Revealing the diversity of Type IIn supernova progenitors through their environments}

\begin{abstract}

Type~IIn~supernovae (SNe~IIn) are hydrogen-rich explosions embedded in dense circumstellar medium (CSM), which gives rise to their characteristic narrow hydrogen emission lines. 
The nature of their progenitors and pre-explosion mass loss remains, however, poorly understood. 
This study analyzes the local stellar environments of a volume‑limited sample ($z < 0.02$) of 31 SNe~IIn using high-resolution Hubble Space Telescope (HST) imaging. 
The environments are found to be highly diverse and are classified into three categories: inside star‑forming regions (\textit{Class~1}), outside but near star‑forming regions (\textit{Class~2}), and regions with no obvious star formation (\textit{Class~3}).
Bright SNe~IIn ($M_{\rm peak}<-19.5$~mag) predominantly occur in Class~1 environments, supporting the origin of very massive progenitors, while the faint SNe~IIn ($M_{\rm peak}<-15.5$~mag) are associated with Classes~2 and 3 environments, consistent with the less massive progenitors.
SNe~IIn with intermediate peak magnitudes ($-19.5<M_{\rm peak}<-15.5$~mag) appear in all three types of environments.
Furthermore, directly detected SN~IIn progenitors are systematically brighter and/or bluer than the youngest stellar populations in their environments, implying a non‑quiescent pre‑explosion state or past binary interactions.
These results point to multiple progenitor channels of SNe~IIn, spanning a wide range of masses, evolutionary stages, and potential binary interaction histories.

\end{abstract}

\section{Introduction} \label{sec:intro}

Type~IIn~supernovae (SNe~IIn), characterized by narrow hydrogen (H) emission lines in their spectra \citep{1990Schlegel}, constitute a fascinating yet enigmatic class of stellar explosions. 
Their defining observational features are attributed to the strong interaction between the expanding ejecta and a dense, pre-existing circumstellar medium (CSM). 
This interaction gives rise to a remarkable diversity in observational behaviors (e.g., \citealp{2012Kiewe,2013Taddia,2020Nyholm}).
Many events, such as SN 1988Z and SN 2005ip,  display luminous, slowly evolving light curves that can last for years, accompanied by persistent interaction‑dominated spectral signatures (e.g., \citealp{2012Stritzinger,2017Smith_1}). 
Rapidly evolving events like SN 1994W and SN 2009kn exhibit a light‑curve plateau followed by a sharp decline and narrow P Cygni profiles (e.g., \citealp{2013Mauerhan}).
Objects similar to SN 2009ip are notable for their pre‑explosion activity, especially the precursor outburst roughly one month before a second, more luminous event (e.g., \citealp{2018Pastorello}).
Transitional IIn SNe, such as SN 1998S and SN 2013cu, initially show brief signatures of CSM interaction. These spectral features typically fade within weeks, after which their spectra come to resemble those of Type IIL or IIb SNe(e.g., \citealp{2015Shivvers}).
Some superluminous events like SN 2006gy are spectroscopically Type IIn SNe, with CSM interaction invoked as the primary power source (e.g., \citealp{2007Smith}). 
Additional, a few Type Ia SNe exploding within dense, H-rich CSM are classified as Ia‑CSM, sharing spectroscopic similarities with SNe IIn (e.g., \citealp{2015Fox_1}). 
The evolving understanding is that SNe IIn represent a heterogeneous class, likely arising from multiple progenitor channels and mass-loss histories.

The origin of the dense CSM, which typically requires enhanced mass loss ($\dot{M}>10^{-3}\ {\rm M_{\odot}\ yr}^{-1}$) in years to thousands of years preceding core collapse \citep{2014Moriya}, remains a key question in stellar astrophysics. 
Several mass-loss processes have been investigated. Steady line-driven winds, closely tied to stellar mass and metallicity \citep{1999Lamers,2001Vink}, typically yield mass-loss rates on the order of $10^{-6}$--$10^{-5}$~$M_{\odot}\ {\rm yr}^{-1}$.
The mass loss during the evolved phase of massive RSGs (which are known Type II-P SN progenitors; \citealt{2009Smartt}) can be stronger but still uncertain by orders of magnitudes (e.g., \citealp{1988deJager,2020Beasor,2023Yang,2025Antoniadis}). Episodic mass loss can be triggered by bi-stability jump \citep{1995Lamers,1999Vink}, pulsations \citep{2010Yoon}, S Doradus-type instabilities \citep{2025Levesque}, or wave-heating energy depositions \citep{2012Quataert,2014Shiode}. Such episodic eruptions could naturally produce multiple, dense circumstellar shells. Collisions between dense CSM shells are considered to explain the double‑peaked light curves of SN 2009ip‑like transients \citep{2014Graham,2015Martin}, as well as the SN 1994W-like events \citep{2016Dessart}.
Through pulsations and S Doradus-type instabilities, mass loss rates are generally on the order of $\le10^{-4}$~$M_{\odot}\ {\rm yr}^{-1}$. The efficiency of the wave-heating mechanism remains controversial \citep{2021Wu,2021Leung}, with suggestions that its main outcome may be envelope inflation rather than substantial mass ejection \citep{2014Mcley,2017Fuller,2022Wu}. 
Moreover, binary interaction such as Roche-lobe overflow, common-envelope ejection, or stellar mergers, has been widely proposed as an efficient channel for pre-SN mass loss \citep{2014Smith,2020Schroder,2024Ercolino}.

Direct detections for progenitors of clearly confirmed SNe~IIn in pre-explosion images have been limited. In the few cases where progenitors have been identified, they are consistently very luminous and blue, often exhibiting significant photometric variability, which are commonly reminiscent of luminous blue variable stars \citep[LBVs;][]{2007Gal-Yam,2011Smith}. 
In the standard picture of single star evolution, LBVs are considered as a short-lived transitional phase for the most massive stars ($M_{\rm ini}>30$--40$M_{\odot}$) on their way to becoming Wolf–Rayet (WR) stars. The dense CSM observed for SNe~IIn can thus be naturally interpreted by the enhanced wind and/or giant outburst similar to those seen from local LBVs \citep{1994Humphreys,1999Humphreys}. 
However, the direct explosion of LBVs is inconsistent with the standard picture of massive single-star evolution (the Conti scenario, \citealt{1971Conti}; see also \citealt{2003Heger}), and it is unclear whether all SNe~IIn progenitors, especially whose without direct detections, resemble LBVs prior to the explosion. Alternatively, the high luminosity of the detected progenitors could be attributed to pre-SN outbursts \citep[e.g.,][]{2024Cheng}, or the uncertain evolutionary pathways such as binary mergers \citep[e.g.,][]{2014Justham}.

Another powerful approach to constraining progenitors of core-collapse SNe is to analyze their local stellar environments. The principle is that massive stars ($M_{\rm ini}>8~M_{\odot}$) form in groups and share similar ages and metallicities, and they have short lifetimes and remain close to their birth sites. Since this technique does not rely on serendipitous pre-explosion images, it significantly enlarges the sample of SNe~IIn available for analysis. It also helps to avoid biases introduced by precursor outbursts \citep{2014Ofek_1}, the self-obscuration of LBVs \citep{2010Wachter}, and the potentially wide range of intrinsic luminosities among progenitors. Moreover, the physics of the evolution of most surrounding stars, in particular those on the main sequence is generally better understood than the nature of the presumed LBV progenitors. Using pixel statistics techniques, SNe~IIn have been found to have a similar association with star formation
to the overall SNe~II(P) population and red supergiants (RSGs), which are weaker than that of Type~Ic~SNe and LBVs \citep{2008Anderson,2012Anderson,2014Habergham,2017Kangas}. Studies based on spatially resolved spectroscopy yield similar results, showing comparable environments for SNe~IIn and SNe~II in terms of metallicity and H$\alpha$ equivalent width \citep{2025Pessi,2025Xi}. These findings challenge the conventional view that SNe~IIn originate exclusively from very massive progenitors.

Furthermore, a key insight emerges from recent works that the SNe~IIn environments are very diverse: some are linked to young stellar populations, tracing the high-mass progenitors, while others associated with old stellar populations, indicative of moderately massive progenitors \citep{2018Galbany,2022Ransome,2023Xiao}. Particularly, some SNe~IIn, such as SN~2010jl, align well with very young star-forming regions \citep{2024Niu}, while others like SN~2009ip are located in a very sparse and old environments \citep{2016Smith}, which is inconsistent with the LBV progenitor of 50--80~$M_{\odot}$ as inferred from the direct detection (see also \citep{2022Smith} for a possible obscured young stellar cluster). These works come into a growing evidences pointing toward significant diversity among SNe~IIn progenitors.

Most previous studies relied on comparative analyses between different types of SNe and massive stars, rather than directly characterizing the stellar populations surrounding SNe~IIn. Recent work by \citet{2023Moriya} has explored the environmental dependence of SNe~IIn using the integral field spectroscopy of their host galaxies, finding tentative correlations between SN properties and local interstellar conditions like metallicity and specific star formation rate. On the other hand, high-resolution imaging from the Hubble Space Telescope (HST) enables direct measurements of stars in the SN vicinity, providing powerful constraints on their ages and masses \citep{2018Maund,2023Sun_2}. However, systematic studies of SNe IIn environments utilizing HST data as well as comparisons between the detected progenitors and their local stellar populations remain limited.  

In this paper, we present a systematic study of local environments of 31 SNe~IIn. We utilize a volume-limited sample ($z < 0.02$) observed with high-resolution HST imaging to ensure that the physical scales of star-forming complexes, crucial for understanding the local environments, are spatially resolved.
We focus on the diversity of their environments, and its connections to SN properties, as well as comparisons between the progenitors and environmental stellar populations. 
This paper is structured as follows. Section~\ref{sec:data} describes the sample selection and data reduction. We present the three classes of SNe~IIn local environments in Section~\ref{sec:3env}. In Section~\ref{sec:env_peak}, we map the environmental dependence to infer possible progenitor channels and mass-loss mechanisms for SNe~IIn of different luminosities. In Section~\ref{sec:env_prog}, we infer the pre-explosion state of the progenitors by comparing them with local stellar populations. We summarize our conclusions in Section \ref{sec:summary}.

\section{Data}\label{sec:data}

\subsection{SNe IIn sample}\label{sec:IInsample}

We started with the SNe~IIn compilation provided by \citet{2014Habergham,2021Ransome,2022Ransome,2025Ransome,2024Hiramatsu}. While discrepancies exist among these catalogs due to the complex nature of Type~IIn SNe and heterogeneous community data sources, they encompass nearly all documented SNe~IIn by the end of 2023.
A few spectroscopically confirmed SNe~IIn from individual studies were excluded in the above references. These objects are included in our work (SN~2016jbu, \citealp{2022Brennan_1}; SN~2010bt,
\citealp{2018EliasRosa}; SN~2010jp, \citealp{2012Smith}). 
The classification of SN~1961V has been historically debated, between a true SN~IIn and an SN impostor (i.e. non-terminal explosions of massive stars; \citealp{2004Chu,2011Kochanek,2012VanDyk}). Recent evidence supports the core‑collapse interpretation \citep{2019Patton}, and we therefore retain it in our sample.
Some intermediate-luminosity optical transients such as SN~2008S and NGC~300OT, while showing narrow emission lines akin to SNe~IIn, are attributed to distinct phenomena like SN~impostors, electron capture supernovae, or intermediate-luminosity red transients \citep{2009Berger,2011Smith4,2021Cai} and are thus excluded from this work.

We further required $z<0.02$, where stellar populations in the SN environments are spatially resolved in HST high-resolution imaging. Following an HST archive search (query date 2025-06-26), we retained 31 SNe~IIn that have been observed by at least one broadband filter with either the Ultraviolet-Visible channel (UVIS) of Wide Field Camera 3 (WFC3) or the Wide Field Channel (WFC) of Advanced Camera for Surveys (ACS). 

For SNe within 10~Mpc, Cepheid distances of their host galaxy are adopted when available (SNe~1961V, 1997bs, 1978K, \citealp{2011Kochanek,2012VanDyk,2015Adams,2020Chiba}). We adopted Hubble distances assuming $H_0 = 73.3~\mathrm{km~s^{-1}~Mpc^{-1}}$ \citep{2022Riess} for other SNe. 
All distances for the whole sample are listed in Table~\ref{tab:result}.

\subsection{HST photometry}\label{sec:hstphot}

Flat-fielded files (\texttt{*$\_$flc.fits} or \texttt{*$\_$flt.fits}) were retrieved via the Mikulski Archive for Space Telescopes (MAST)\footnote{\url{https://mast.stsci.edu}}, as listed in Table~\ref{tab:HSTtab}. 
Observations with exposures at only a single point (i.e. not taken with a series of offsets) were excluded due to inadequate cosmic-ray removal.
The dithered exposures were combined using the \textsc{astrodrizzle} package to remove cosmic rays and create drizzled science images, which served as reference frames for subsequent photometry.
Point-spread-function (PSF) photometry was performed with the \textsc{dolphot} package\footnote{\url{http://americano.dolphinsim.com/dolphot/}} \citep{dolphot.ref,Dolphin.ref2}. The parameters \texttt{FitSky=2} and \texttt{RAper = 3} were applied. All other parameters followed recommendations in the \textsc{dolphot} User's Guide.

We applied the following criteria in each filter to identify resolved stellar sources:

 (1) \texttt{object type} = 1, to select stellar objects;

 (2) \texttt{signal-to-noise ratio} $\ge$ 5;

 (3) $-$0.5 $\le$ \texttt{sharpness} $\le$ 0.5, to select point-like sources;

 (4) \texttt{crowding} $\le$ 2, to exclude overly crowded and contaminated measurements;

 (5) \texttt{quality flag} $\le$ 2, to ensure successful photometric fitting.

Artificial stars were randomly inserted to quantify and account for additional photometric uncertainties arising from source crowding and imperfect sky subtraction.

\subsection{Pinpointing the SN locations}\label{sec:pin}

The SN positions on the HST images were carefully determined.
Several SNe have had their explosion sites or progenitors identified within HST imaging, including SN~1961V \citep{2011Kochanek,2019Patton}, SN~1997bs \citep{1999VanDyk,2000VanDyk}, SN~1998S \citep{2012Mauerhan}, SN~1999el \citep{2002Li}, SN~2005gl \citep{2007Gal-Yam}, SN~2005ip \citep{2020Fox}, SN~2006gy \citep{2015Fox}, SN~2009ip \citep{2022Smith,2010Smith}, SN~2010bt \citep{2018EliasRosa}, SN~2010jl \citep{2011Smith,2024Niu}, SN~2015bh \citep{2016EliasRosa}, SN~2016jbu \citep{2018Kilpatrick}, and SN~2021adxl \citep{2024Brennan}. These HST images served as our astrometric reference frames.
For SN~1978K \citep{1993Ryder}, SN~2010jj \citep{2021Schulze}, SN~2013gc \citep{2019Reguitti}, and SN~2015bf \citep{2021Lin}, post-explosion imaging obtained from ground-based telescopes were taken to determine the SN site. Astrometric transformations were performed as detailed in \citet{2025Niu}. 
The uncertainties of transformation were typically $<1$ pixels for HST-to-HST alignments and 2--4 pixels for for ground-based-telescope-to-HST alignments. At the distances of our SN sample, 1 HST pixel corresponds to 1--15~pc. For SN~2017hcc, it was still bright on the HST images analyzed in this work, from which we directly obtained the SN position. For all other SNe, theirs WCS were validated using the python package \textsc{twirl} \citep{2010Lang,2022Garcia} with \textit{Gaia} reference stars. The resulting uncertainties were comparable to those from ground-based-telescope-to-HST transformations.

\section{Three classes of SN environments}\label{sec:3env}

For each SN, a stellar surface density map was constructed within an 800 $\times$ 800~pc region centered on the SN position. The map was generated by binning the resolved point sources into a two‑dimensional histogram and convolving it with a Gaussian kernel to obtain a smoothed density map ($Z_{\rm smooth}$). 
The kernel width represents the spatial resolution of the density map; a larger width suppresses noise at the expense of spatial details, while an overly small width introduces noise. 
Because the SNe in our sample span a range of distances and have been observed with different HST bands and exposure times, we adopted an empirical approach to set the width \citep{2017Gouliermis,2017Sun}.
After testing a range of kernel sizes, we found that using the median distance to the 5th-nearest neighbor as the width provides an optimal balance between resolution and noise. The adopted width typically lie between 30 and 100~pc. 

\begin{figure*}[htbp] 
    \centering %
    \includegraphics[width=\linewidth]{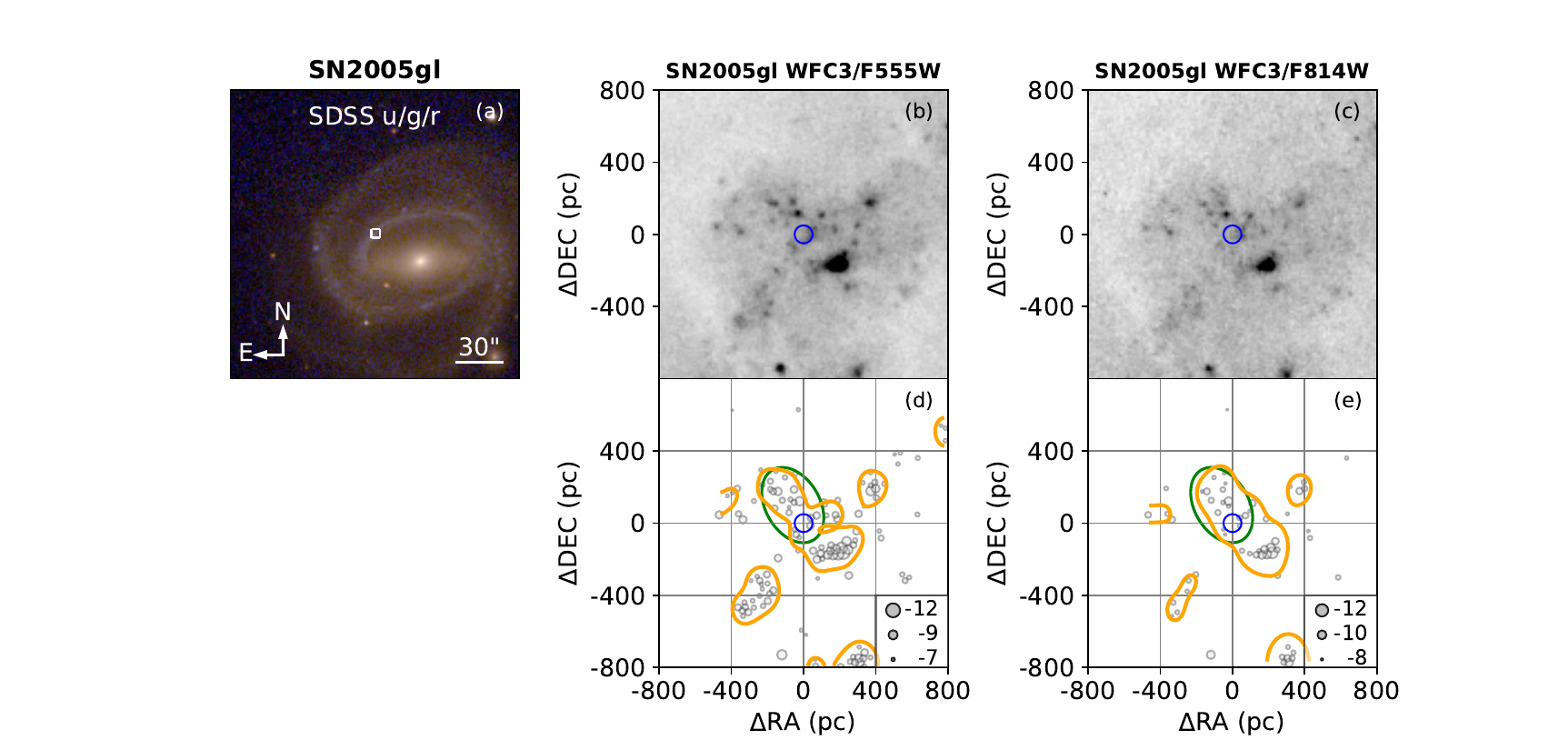}
\caption{The environment of SN\,2005gl is an example of a Class~1 environment, locating within a star-forming region. All panels are oriented with north up and east to the left.
(a) A three-color SDSS $u/g/r$ composite image of the host galaxy, centered on the SN. The white square outlines an 800~pc $\times$ 800~pc region, which is shown zoomed in the subsequent panels. 
(b, c) HST images of the SN site, centered on the SN. (d, e) Spatial distribution of stars detected in the HST images with symbol sizes representing absolute magnitudes corrected for Galactic extinction. 
The orange contour marks a 3$\sigma$ density enhancement above the local background. 
Blue circles in panels (b)--(e) indicate a 50~pc radius centered on the derived SN site, accounting for the positional uncertainties of the SN in Section~\ref{sec:pin}. 
The green elongated circles in panels (d) and (e) outline the nearest stellar overdensity region to the SN, which is spatially distinct from the dense region southeast of the SN. Stars within this circle are considered to be coeval with the SN progenitor and are used for comparison in analysis in Section \ref{sec:env_prog}.
\label{fig:class1_05gl}}
\end{figure*}

\begin{figure*}[htbp] 
    \centering %
    \includegraphics[width=\linewidth]{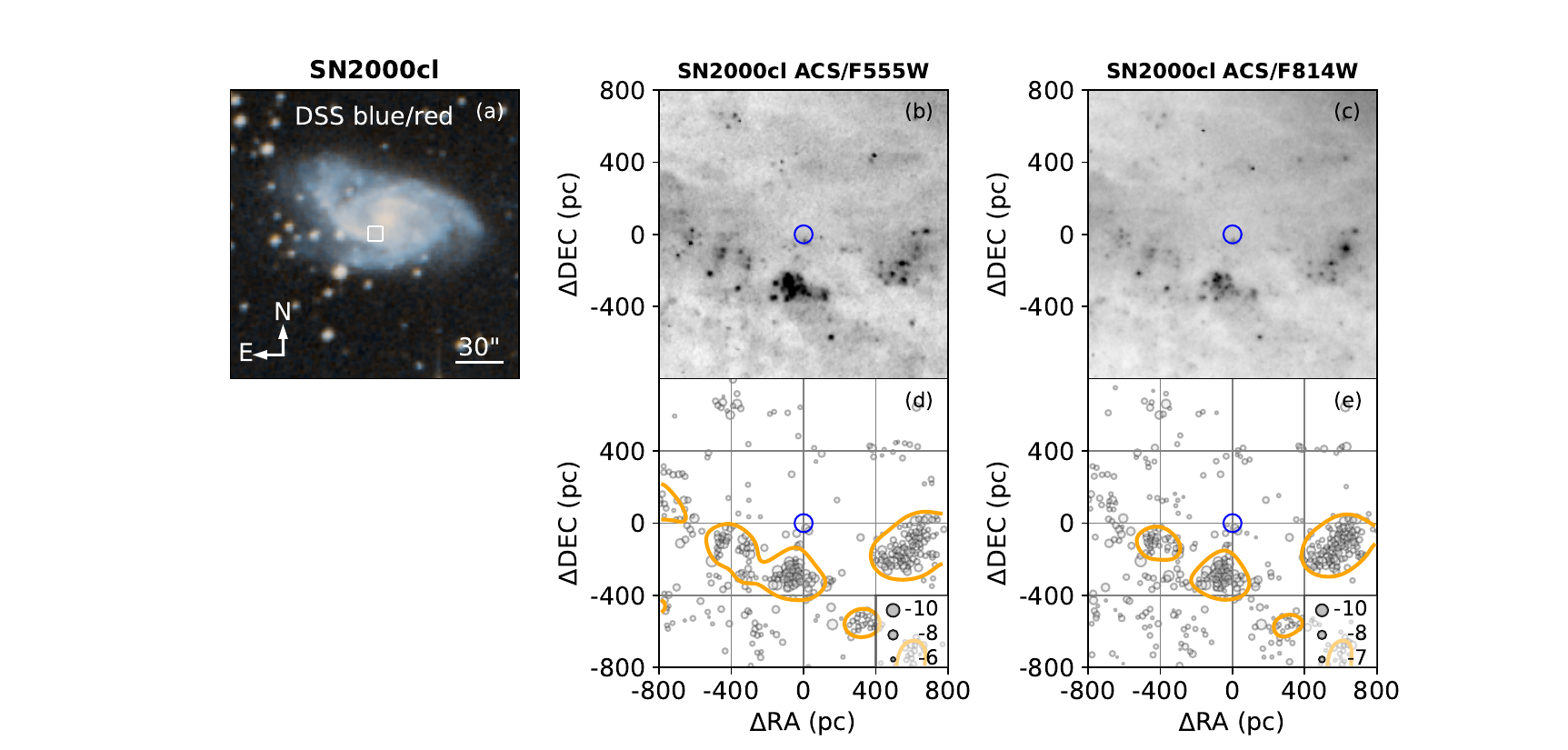}
\caption{Similar to Figure~\ref{fig:class1_05gl}, but for SN\,2000cl. No stellar overdensity is detected within 50~pc of the SN; however, a significant overdensity of stars is detected within 300~pc. This distance approximately corresponds to the spatial displacement within which 90$\%$ of progenitors that have undergone binary interactions. \citep{2025Wagg,2019Renzo}. 
Environments such as that of SN\,2000cl are defined as Class~2.
\label{fig:class2_00cl}}
\end{figure*}

\begin{figure*}[htbp] 
    \centering %
    \includegraphics[width=\linewidth]{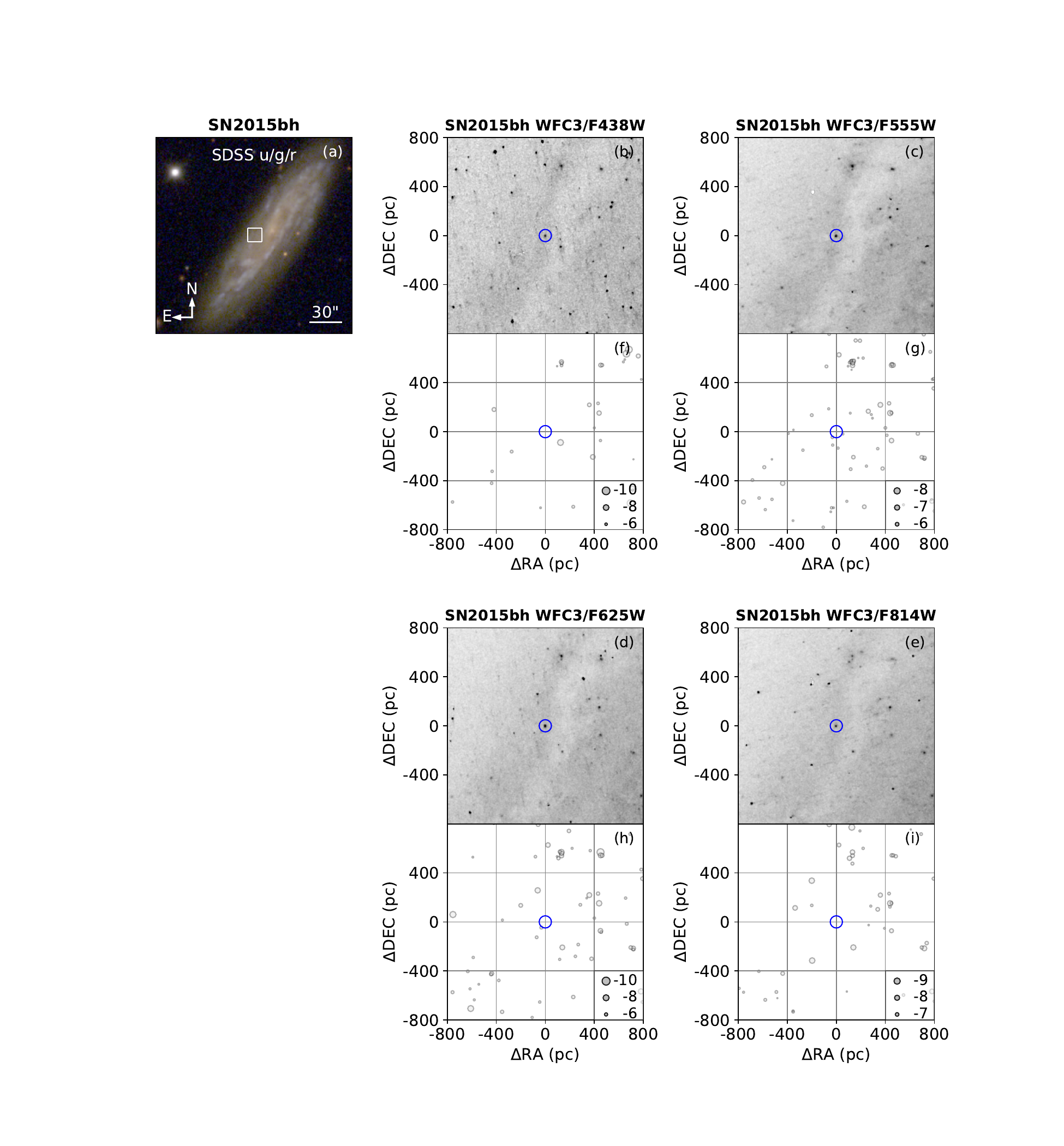}
\caption{Similar to Figures~\ref{fig:class1_05gl} and \ref{fig:class2_00cl}, but for SN\,2015bh. This environment lacks strong evidence for star-forming regions within 300~pc and is therefore defined as Class~3.
\label{fig:class3_15bh}}
\end{figure*}

From a smoothed density map, the 3$\sigma$ density enhancement contour corresponds to pixels with:
\begin{equation}
    Z_{\rm smooth}>m+3\sigma
\end{equation}
where the median $m$ of the map was calculated via sigma-clipped statistics. Figures~\ref{fig:class1_05gl}, \ref{fig:class2_00cl}, and \ref{fig:class3_15bh} demonstrate representatives of SNe~IIn environments in our sample. Orange iso-density contours mark the 3$\sigma$ density enhancements, and these overdensities trace star-forming regions that have not been dissipated. Figures for the whole SNe~IIn sample are in Appendix~\ref{sec:env_appendix}. 

We determined the projected distance to the nearest 3$\sigma$ contour as the Euclidean distance from the SN position to the nearest point on any $3\sigma$ contour (i.e., the edge of the nearest star-forming regions). If the SN position fell inside a closed contour, the projected distance was set to zero. 

Based on the local surface density map and the measured distances, the 31 SNe IIn were divided into three classes:

\textit{Class~1} : The projected distance is less than 50~pc. The threshold of 50~pc accounts for the typical astrometric uncertainties and the smoothing introduced by the kernel. SNe of this class are likely to have very young and massive progenitors, although chance alignments cannot be entirely ruled out. An example is shown in Figure~\ref{fig:class1_05gl}.

\textit{Class~2}: The projected distance is between 50 and 300~pc. The upper limit of 300~pc corresponds to the projected distance within which approximately 90$\%$ of massive runaway stars from binary systems are predicted to be found \citep{2019Renzo, 2025Wagg}.
The Class~2 events may originate either from older stellar environments or from massive stars that have migrated from the adjacent star-forming complex. A representative case is displayed in Figure~\ref{fig:class2_00cl}.

\textit{Class~3}: The projected distance is greater than 300~pc. These SNe are located far from any significant stellar overdensity and most probably arise from an older, more dispersed stellar population. An illustration is given in Figure~\ref{fig:class3_15bh}.

The interpretations above rely on the assumption that the age of the local stellar population is representative of the progenitor age. This assumption is generally valid but may be less robust in mixed-age or low-density environments. 
To mitigate the potential impact of stellar migration, we adopt a relatively large criterion of 300~pc, which is expected to encompass the majority of migrated progenitors. Unless the migration is extreme, the results should therefore remain robust.
While the environmental age does not provide a precise constraint on individual progenitor mass, it remains a useful statistical indicator for distinguishing between broadly young and old progenitor populations.

Special considerations were taken for SNe~2009ip, 2010jp, 2015bf, 2017hcc, and Gaia14ahl. \citet{2022Smith} points that the nearly constant UV flux at the SN~2009ip site after a decade indicates a possible underlying young star cluster ($<10$~Myr) behind the SN. \citet{2022Corgan} reports an likely extended H$\alpha$ emission behind the SN~2010jp, and they propose that it is related to a recent star formation with at least one late O-type star. 
Based on these indicators of recent star-forming activity, we reclassify these two SNe into Class~1, even though their environments appear sparse in HST resolved stellar photometry.
SN~2017hcc remained significantly bright in 2021; the radius of the region that exceeds the background brightness by 5$\sigma$ is approximately 0.14", corresponding to an obscured region with a radius of 48~pc at the SN distance. Due to the limited depth of the observations of the region hosting SN\,2017hcc (see Appendix Figure~\ref{sec:comp}), we cannot rule out the presence of a compact obscured star-forming region.
A noticeable enhancement in surface brightness is observed for SN~2015bf and Gaia14ahl, yet fewer than 10 resolved stars were detected within 800~pc of these SNe. We suggest that this is likely due to the very shallow detection limits of the observations. Thus, these were reclassified into Class~1. Although the possibility of Class~2 remains speculative, this would not affect our main conclusions, as discussed in the following sections.

While the above reconsideration aims to provide a more physically complete picture of the local environment of SNe~IIn in our sample, it introduces a potential inhomogeneity with respect to the remainder of the sample, which is classified based on the HST stellar density maps.  We note that similar situations could, in principle, affect other SNe~IIn in our sample. For example, some apparently sparse environments might host obscured star-forming regions that are not detected in the available optical HST images due to dust extinction or insufficient depth. Our HST-based classification should be regarded as a conservative estimate that may underestimate the presence of young stellar populations. 

The classifications for the whole sample are listed in Table \ref{tab:result}, with 16 SNe belonging to Class~1, 7 to Class~2, and 8 to Class~3. 
It is important to note that while our sample is volume-limited, historical discovery surveys likely prioritized brighter, star-forming host galaxies, potentially biasing against SNe IIn in quiescent regions \citep{2025Xi}. This may result in a relative undercount of Class 3 events, a limitation that should be interpreted cautiously.
Nevertheless, the presence of SNe IIn across all three environmental classes clearly highlights their diversity, in agreement with previous findings (e.g., \citealp{2022Ransome}).
This diversity argues against very massive stars as the sole progenitor channel for SNe~IIn (if that were the case, all SNe~IIn would be of Class~1, or even Class~2 if progenitors have migrated). This suggests that SNe~IIn progenitors should have multiple evolutionary pathways, with a range of initial masses and lifetimes.

\section{Relation with the SN peak magnitudes}\label{sec:env_peak}

In this section, we explore links between the properties of SN explosions and their environments. First of all, we need to identify a key parameter to characterize the SN explosion. CSM interaction is the dominating power source of SN~IIn light curves; the complex nature of CSM introduces a considerable diversity, including a wide range of peak magnitudes, lasting timescale, rise/decline rates, and short-lived features.
\citet{2024Hiramatsu} measured light curves of a large SNe~IIn sample; they find a tight correlation between rise/decline times and peak magnitude and identify two subgroups of luminous-slow and faint-fast events. They suspect that the luminous-slow subgroup likely originates from more massive progenitors while the faint-fast subgroup may correspond to explosions of less massive progenitors. 
In parallel, \citet{2025Ransome} applies light curve fitting of ejecta–CSM interaction models and reports that SNe~IIn with more massive CSM have longer $r$-band rise/decline times and brighter peak magnitudes (see also \citealp{2020Nyholm}). Therefore it seems that the peak magnitudes of SNe~IIn can serve as a useful proxy of CSM properties. Moreover, the measurement of peak magnitudes is much easier than that of other SNe observables (such as the rise/decline time, which requires good sampling of the light curve, but in our sample such light curves are always not available). In the following subsections, we try to investigate the relation between the peak magnitude and the local environment of SNe~IIn.

Peak magnitudes for the SNe in the sample were collected, and reported host extinctions from the literature were applied. These values, along with corresponding references, are summarized in Table~\ref{tab:result}. We preferentially used $r/R$/unfiltered (unf)-band photometry near the peak. For the 6 objects lacking $r/R$/unfiltered (unf)-band photometry, $B/V/I$-band photometry were used instead, which are very close to $r/R$/unfiltered (unf)-band photometry as optical colors at the SN peak are close to zero \citep{2019Jaeger}. For 10 SNe without photometric observations near the maximum luminosity, the brightest available measurement was adopted as a lower limit of the peak magnitude. 
In the case of SN~2009ip-like objects showing double peaks, the brighter peak of the second event was used in the following analysis. 
After applying the respective distances applied and corrections for both Galactic \citep{2011sfd} and host extinction (from the literature), the resulting peak absolute magnitudes wre plotted ($M_{\rm peak}$) against the SNe environment classes in Figure~\ref{fig:env_mpeak}. 
Within each environment class, the SNe are ordered by their $M_{\rm peak}$ and labelled accordingly. 

\subsection{``Bright'' SNe~IIn}

\begin{figure*}[htbp] 
    \centering %
    \includegraphics[width=0.9\linewidth]{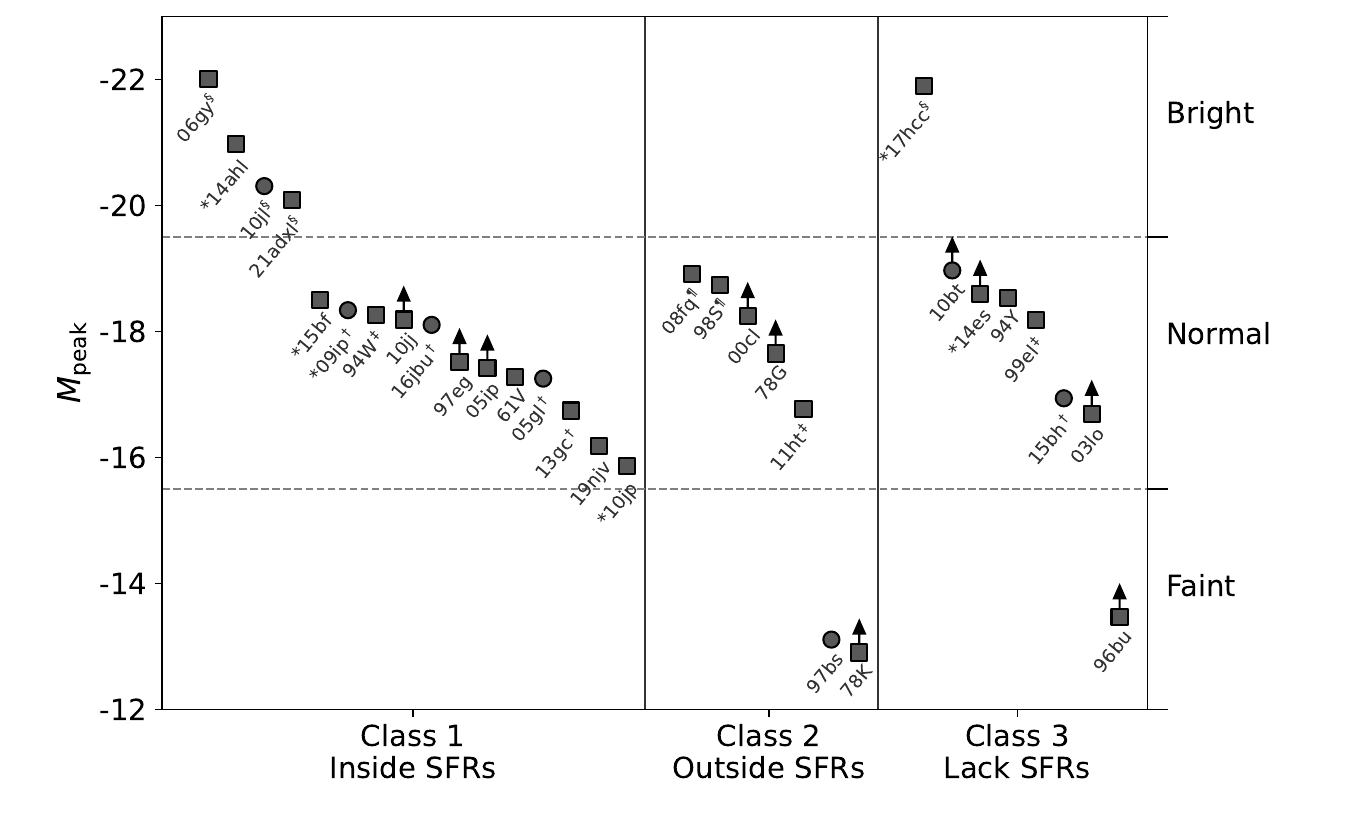}
\caption{Peak absolute magnitude versus local environment class for 31 SNe~IIn at $z < 0.02$ with HST high-resolution imaging.
The local environments of SNe are classified into three types based on resolved stellar populations: inside star-forming regions, outside star-forming regions, or regions lacking obvious star-forming regions. 
Those SNe for which the classification of their host environment is still preliminary are indicted with $*$.
We classify their peak absolute magnitudes as: ``Bright" ($M_{\rm peak}<-19.5$), ``Normal" ($-19.5<M_{\rm peak}<-15.5$), or ``Faint" ($M_{\rm peak}>-15.5$).
Among the SNe IIn, those resembling SN 2009ip, SN 1994W, SN 1998S, and those identified as SLSNe are marked with $\dagger$, $\ddagger$, $\P$ and $\S$, respectively.
SNe whose progenitors have been directly detected are indicated with circles; their progenitors and environmental stars are analyzed in detail in Section \ref{sec:env_prog}.
}
\label{fig:env_mpeak}
\end{figure*}

As shown in Figure~\ref{fig:env_mpeak}, the SNe~IIn in our sample exhibit clear separations in peak magnitude. Five objects with $M_{\rm peak}<-19.5$~mag are notably more luminous than the rest. These are designated as the ``Bright'' SNe~IIn, and some have been identified as superluminous SNe~IIn in literature. Such high luminosities are commonly explained by extreme CSM interaction in iron core-collapse SNe or pair-instability SNe \citep{2007Woosley,2007Smith}, in which a substantial fraction of ejecta kinetic energy is dissipated and efficiently converted to optical radiation \citep{2010vanMarle,2020Suzuki}. To power these bright transients, the CSM mass is estimated to be least 5--10~$M_{\odot}$, and in some cases as high as 20--40~$M_{\odot}$ \citep{2008Smith,2020Nicholl}, that, in turn, implies very massive progenitors.

An alternative model for such luminous event invokes thermonuclear explosions of white dwarfs in binary systems following a common-envelope phase shortly before the Type~Ia~SNe \citep{2021Ablimit}, normally named as Type~Ia-CSM \citep{2003Hamuy,2013Silverman}.  
\citet{2015Leloudas} demonstrated that strong CSM interaction can mask an underlying SN~Ia spectrum, and such Ia-CSM events typically exhibit a bright absolute magnitude range $-19.5>M>-21.6$, overlapping the luminosities of our `Bright'' SNe~IIn.
In this scenario, the progenitor is a low-mass star. 
The well-studied ``Bright'' Type~IIn SN~2006gy exemplifies the ongoing debate over the progenitor identity \citep{2020Jerkstrand}.

Insights of progenitor masses could be inferred from environmental analysis. Four of the five ``Bright'' SNe~IIn in our sample are located in the Class~1 environment. This finding supports their association with very massive progenitors, although alternative channels cannot be completely excluded.
In previous works, we have performed detailed modelling of stellar populations in vicinity of SN~2010jl and found that it likely originated from a very young environment with very recent star formation \citep{2024Niu}. Similarly, \citet{2007Ofek} and \citet{2024Brennan} also report star-forming environments for the ``Bright'' Type~IIn~SNe~2006gy and 2021adxl, respectively. While there are no obvious resolved stellar populations near the SN~2017hcc, its environment classification might be affected by the bright SN itself and the shallow detection limit in the UV band, and we can not rule out an underlying star-forming region at the SN position. Therefore, the ``Bright'' SNe~IIn are most likely to originate from very massive progenitors.

\subsection{``Normal'' SNe~IIn}

The majority of SNe~IIn share a typical peak magnitude in the range of $-19.5<M_{\rm peak}<-15.5$~mag and are donated as ``Normal'' SNe~IIn (e.g., \citealp{2014Richardson,2025Pessi}). For their pre-SN mass loss, observational evidences suggest high rate of $\ge10^{-3}M_{\odot}\ {\rm yr}^{-1}$ (e.g. \citealp{2012Kiewe,2014Moriya}). The associated CSM is estimated to have a mass range of 0.5--8~$M_{\odot}$, extending to radii of tens of AU (e.g., \citealp{2025Ransome}). Interestingly, these ``Normal'' SNe~IIn are observed across all three environment classes.  This spread suggests a broad distribution of progenitor ages and initial mass, indicating that their CSM properties are insensitive to the initial mass. This picture challenges scenarios in which a dominant, mass‑dependent process (such as pulsation‑driven superwinds; \citealp{2010Yoon}) solely governs the pre‑explosion mass loss. It is also possible that binary‑mediated mass loss, where system parameters (e.g., mass ratio, orbital separation) play the decisive role, produce the observed diversity without requiring a narrow range of progenitor masses.

\citet{2025Kankare} recently investigated the environments of SN 1994W-like transients, which is a subtype of SNe~IIn with intermediate luminosity. They found that their explosion site resembles that of low-mass RSGs. Our sample includes three such events: SN~1994W, SN~2011ht, and SN~1999el, which are classified here as Class~1, 2, and 3, respectively. The association with Class 2/3 environments supports the scenario of a less massive progenitor. For SN~1994W, our Class~1 designation agrees with their HST-based note of a likely cluster at the SN site.

\subsection{``Faint'' SNe~IIn}\label{sec:faint}

Three SNe~IIn in our sample exhibit peak magnitudes fainter than $M_{\rm peak}<-15.5$~mag, approaching the lower luminosity limit of core-collapse SNe. Two of them (SNe\,1978K and 1996bu) were not observed at the peak light, yet their lower limits of peak brightness are $\sim$2.5~mag fainter than the faintest among ``Normal'' SNe~IIn, justifying their classification as ``Faint''. Their inferred pre-SN mass loss rates are  10$^{-4}$--10$^{-3}$~$M_{\odot}$\ ${\rm yr}^{-1}$ \citep{1993Ryder,1995Chugai,2012Kochanek,2020Chiba,2015Adams}, lower than those of ``Bright" and ``Normal" SNe~IIn.

At such faint magnitudes, the population overlaps with non-terminal SN impostors (e.g., outbursts of $\eta$ Carinae and P Cygni,  \citealp{2011Smith4,2011Smith5}). This raises a question of whether the ``Faint'' SNe~IIn represent genuine terminal explosions or the brightest members of the impostor population.
In particular, SN~1997bs has long been considered as an SN impostor \citep{2000VanDyk,2011Smith4,2012Kochanek}. If this interpretation holds, the ``Faint'' SNe~IIn should be exclusively observed in Classes~1 and 2. However, this contradicts our findings that all ``Faint'' objects are associated with Classes~2 and 3, passively star-forming environments. 

Although the Class 2 association could permit a scenario involving migrated massive progenitor stars, the environmental distribution more strongly supports an alternative scenario that ``Faint'' SNe~IIn were genuine but subluminous SNe~IIn resulting from weak explosions ($<10^{51}$~erg) of stars with masses of 8--10 $M_{\odot}$. In such SNe, very little $^{56}$Ni ($\sim$10$^{-3}\ M_{\odot}$) was synthesized, and conversion from shock energy to radiation energy through CSM interaction is not very efficient.
This scenario is also supported by the observed low ejecta velocities \citep{2000VanDyk,2011Smith5,2012SNimp,1993Ryder}. Moreover, \citet{2015Adams} points out that ``Faint'' Type~IIn~SN~1997bs has continued to fade in optical and infrared bands and remains significantly fainter than its progenitor star, a behavior they argue is consistent with a terminal supernova explosion.
It should be noted that the ``Faint'' SNe~IIn discussed here are still considerably brighter than many known SN impostors (peak magnitudes of about $-$10~mag; \citealt{2006Maund,2011Smith4}). Environmental studies of such fainter impostors can be found in \citet{2014Habergham}.

\begin{table*}
\centering
\caption{Metallicity, distance, peak magnitude, host extinction, and environment class for the SN~IIn sample in this work.} \label{tab:result}
\begin{tabular}{lllll}
\hline 
\hline
Name & 12+log([O/H]) & Distance & $M_{\rm peak}$$^{*}$ & $A_{\rm host}$ \\
 & dex & Mpc & mag & mag \\
\hline
\multicolumn{5}{c}{Class\,1} \\
SN1961V & 8.77$^a$ & 9.3 & 12.70(0.00)$^b$ & -  \\
SN1994W & 8.61$^c$ & 16.9 & 13.30(0.00)$^d$ & 0.37$^c$  \\
SN1997eg & 8.53$^e$ & 35.6 & $<$15.27$^f$ & -  \\
SN2005gl & 8.59$^c$ & 63.4 & 17.00(0.10)$^{\dagger,c}$ & 0.09$^g$  \\
SN2005ip & 8.84$^{a,pp}$ & 29.5 & $<$15.04$^c$ & $\approx0$$^{pp}$  \\
SN2006gy & 8.70$^h$ & 78.8 & 14.22(0.03)$^i$ & 1.40$^{i,h}$  \\
SN2009ip & 8.25$^j$ & 24.1 & 13.65(0.10)$^{\dagger,c}$ & 0.03$^k$  \\
SN2010jj & - & 70.6 & $<$16.22$^l$ & -  \\
SN2010jl & 8.25$^m$ & 43.8 & 13.00(0.10)$^m$ & 0.06$^k$  \\
SN2010jp & 8.54$^n$ & 37.7 & 17.20(0.00)$^o$ & $\approx0$$^p$  \\
SN2013gc & - & 13.9 & 14.66(0.14)$^{\dagger,q}$ & $\approx0$$^q$  \\
Gaia14ahl & - & 69.7 & 14.17(0.00)$^k$ & 0.07$^k$  \\
SN2015bf & 8.49$^r$ & 58.3 & 15.66(0.05)$^s$ & 0.19$^s$  \\
SN2016jbu & 8.66$^u$ & 19.6 & 13.80(0.05)$^{\dagger,t}$ & $\approx0$$^u$  \\
SN2019njv & - & 59.8 & 18.08(0.06)$^v$ & -  \\
SN2021adxl & 7.77$^w$ & 73.8 & 14.32(0.02)$^w$ & $\approx0$$^w$  \\
\hline
\multicolumn{5}{c}{Class\,2} \\
SN1978G & 8.19$^r$ & 12.7 & $<$12.90$^b$ & - \\
SN1978K & 8.07$^x$ & 4.61 & $<$16.00$^y$ & 0.23$^z$ \\
SN1997bs & 8.59$^r$ & 9.4 & 16.83(0.03)$^{aa}$ & $\approx0$$^{aa}$ \\
SN1998S & 8.67$^c$ & 12.3 & 12.19(0.00)$^c$ & 0.43$^{bb}$ \\
SN2000cl & 8.53$^r$ & 37.7 & $<$14.80$^{dd}$ & - \\
SN2008fq & 8.58$^r$ & 43.5 & 15.40(0.00)$^{ee}$ & 0.99$^{ff}$ \\
SN2011ht & 8.20$^{gg}$ & 14.7 & 14.15(0.09)$^c$ & 0.06$^k$ \\
\hline
\multicolumn{5}{c}{Class\,3} \\
SN1994Y & 8.74$^{ii}$ & 34.8 & 14.20(0.02)$^{jj}$ & $\approx0$$^c$ \\
SN1996bu & 8.28$^{e}$ & 16.0 & $<$17.60$^{hh}$ & - \\
SN1999el & 8.17$^e$ & 19.2 & 14.52(0.04)$^{cc}$ & 0.47$^{cc}$ \\
SN2003lo & 8.61$^c$ & 57.4 & $<$17.20$^c$ & - \\
SN2010bt & 8.61$^r$ & 66.3 & $<$16.00$^{kk}$ & 0.80$^{kk}$ \\
SN2014es & - & 80.4 & $<$16.00$^{ll}$ & - \\
SN2015bh & 8.46$^{mm}$ & 26.2 & 15.35(0.07)$^{\dagger,nn}$ & 0.15$^k$ \\
SN2017hcc & 8.49$^{oo}$ & 71.0 & 12.46(0.06)$^{oo}$ & 0.03$^k$ \\
\hline
\end{tabular}
\begin{tablenotes} 
\footnotesize 
\item[1] $^{*}$ Peak magnitudes for SN~1978K ($B$), SNe~2010jl and 2013gc ($I$), and SNe~1978G, 1996bu, 1994W ($V$) were obtained in the specified bands; the remaining SNe were observed in $R$, $r$, or unfiltered bands. Galactic extinctions have been corrected based on the SFD extinction map \citep{2011sfd}. \\
\item[3] $^{\dagger}$: SN~2009ip-like, taking the peak magnitude of the second event. \\
\item[4] References: a: \citet{2019Xiao}, b: \citet{1999Barbon}, c: \citet{2015Taddia}, d: \citet{1998Sollerman}, e: \citet{2014Habergham}, f: \citet{2004Tsvetkov}, g: \citet{2007Gal-Yam}, h: \citet{2007Ofek}, i: \citet{2007Smith}, j: \citet{2014Margutti}, k: \citet{2024Bilinski}, l: \citet{2014Ofek_1}, m: \citet{2011Stoll}, n: \citet{2022Corgan}, o: \citet{Maza2010}, p: \citet{2012Smith}, q: \citet{2019Reguitti}, r: \citet{2025Xi}, s: \citet{2021Lin}, t: \citet{2022Brennan_1}, u: \citet{2022Brennan}, v: ZTF, w: \citet{2024Brennan}, x: \citet{2020Chiba}, y: \citet{1993Ryder}, z: \citet{1995Chugai}, aa: \citet{2000VanDyk}, bb: \citet{2000Fassia}, cc: \citet{2002Carlo}, dd: \citet{2001Stathakis}, ee: \citet{2015Bilinski}, ff: \citet{2013Taddia}, gg: \citet{2012Roming}, hh: \citet{1996Armstrong}, ii: \citet{2012Kelly}, jj: \citet{2001Ho}, kk: \citet{2018EliasRosa}, ll: \citet{2014Li}, mm: \citet{2017Thone}, nn: \citet{2016delaRosa}, oo: \citet{2023Moran}, pp: \citet{2012Stritzinger}
\end{tablenotes} 
\end{table*}

\subsection{Significance of the Environment–Luminosity Correlation}

Fisher exact tests were performed to assess the association of ``Bright''/``Faint'' vs. Class~1/non‑Class~1 environments. 
Given that the environmental Class of two bright SN (SN~2017hcc and Gaia14ahl) remains uncertain (see Section~3), we conducted a sensitivity analysis under three scenarios: 
(1) both are assigned to Class~1; 
(2) one belongs to Class~1 and the other belongs to non-Class~1; 
(3) both are assigned to non-Class~1. 
The corresponding $p$ values for these scenarios are 0.0048, 0.0048, and 0.214, respectively. 
The association is statistically significant ($p<0.05$) under the first two scenarios, which we consider the most realistic given the potential observational limitations (e.g.,  obscuration by the SN itself or shallow UV imaging depth). Only under the most conservative case, the correlation fail to meet the conventional significance threshold.

The observed absence of ``Faint'' SNe~IIn in Class~1 is unlikely because of the observational bias. SNe are usually discovered through image subtraction techniques, which are sensitive to detecting faint transients even in bright regions such as star-forming complexes. This view is supported by \citet{2025Senzel}, which shows that the fainter SNe~Ia are actually more frequent in regions with higher surface brightness. Thus, the exclusive association of ``Faint'' SNe~IIn with older stellar populations appears robust.

Our results align with the environmental trends reported by \citet{2023Moriya}, who found that SNe~IIn with higher peak luminosities tend to occur in environments with lower metallicity and/or younger stellar populations, with correlation coefficients of approximately $p=0.6\pm0.1$ and $p=0.4\pm0.1$, respectively. 
In particular, their results reinforce our conclusion for the ``Bright'' SNe~IIn; their brightest SNe~IIn of $M\leq -19$ (corresponding to our ``Bright'' subsample) are clearly associated with younger stellar populations, while within their normal-luminous SNe~IIn ($M\geq -19$), no strong trend is found between peak magnitude and stellar population age or metallicity.
The agreement between our independent samples and methodologies strengthens the robustness of the emerging picture regarding the diversity of SNe~IIn progenitors and the interpretation that ``Bright'' SNe~IIn are statistically linked to young star‑forming regions (Class~1), while ``Faint'' SNe~IIn exclusively arise in older, non‑star‑forming environments (Classes~2 and 3).
We further note that the sample of \citet{2023Moriya} does not include counterparts to our ``Faint'' SNe~IIn; our work therefore extends the environmental mapping of SNe~IIn to lower luminosities by revealing their exclusive association with much older stellar environments.

Our reclassification for the environmental Class for a few SNe could, in principle, work for other SNe~IIn in our sample. For example, some sparse environments might host undetected star-forming regions due to dust extinction or insufficient imaging depth. 
Nevertheless, such potential biases would not alter our main conclusions: 
(1) For the ``Bright'' SNe~IIn, we have considered the potential influence on the statistical significant above, and the luminosity-environment pattern is further supported by the independent findings of \citet{2023Moriya}; 
(2) For the ``Normal'' SNe~IIn, our main conclusion is the diversity across all three environments. This result is not sensitive to the possible presence of a few undetected star-forming regions;  
(3) For the ``Faint'' SNe~IIn, the three defining objects are among the photometrically best‑observed in our sample (see Appendix \ref{sec:comp}), making their exclusive link to Class~3 environments the least likely to be affected by detection limits.
While future deeper, multi-wavelength observations may refine the environmental census, the luminosity‑environment correlation and the diversity of environments reported in this work are robust against the potential biases discussed here.

\begin{figure}[htbp] 
    \includegraphics[width=0.85\linewidth]{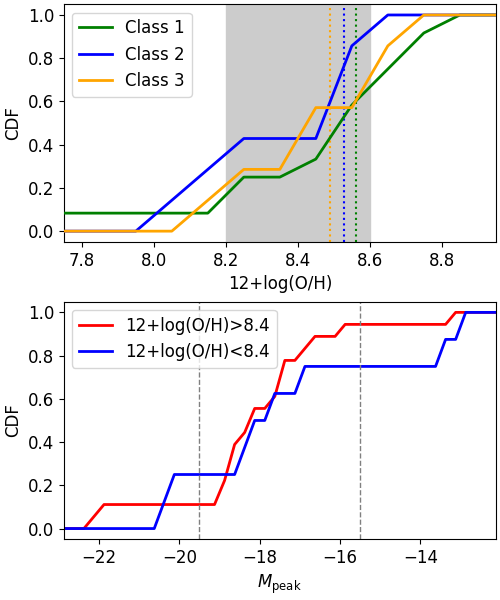}
\caption{Top panel: Cumulative distribution functions (CDFs) of metallicity for SNe IIn in three classes of environments. The three dotted vertical line indicate the median metallicities of each class. The shaded region marks the averaged metallicity and its 1$\sigma$ uncertainty (8.4\,$\pm$\,0.2 dex) of SNe IIn from \citet{2025Xi}. Bottom panel: CDFs of peak absolute magnitude for metal-rich and metal-poor SNe IIn. The gray vertical lines mark classifications of the absolute magnitudes. }
\label{fig:metallicity}
\end{figure}

\subsection{Potential Role of Metallicity}

Metallicity is a key parameter influencing stellar evolution, particularly through line-driven wind mass loss \citep{2001Vink,2008Puls} and associated phenomena such as the bi-stability jump \citep{1995Lamers,1999Vink}. It has therefore been proposed as a factor that could shape the CSM and, in turn, affect the observed properties of SNe IIn (e.g., \citealt{2015Taddia}).

To assess whether metallicity significantly modulates the environment–luminosity trend presented in Figure 4, we performed two tests using the explosion-site metallicities listed in Table 1. The metallicity distributions across our three environment classes are shown in the top panel of Figure~\ref{fig:metallicity}. Although the mean metallicity decreases marginally from Class 1 to Class 3, a statistical test indicates that this weak trend has a 43\% probability of arising from random sampling drawn from the same reference distribution. All three averaged metallicities are consistent with the mean metallicity of SNe IIn reported by \citet{2025Xi} within its uncertainty, suggesting no pronounced difference in metallicity. 

The suggested tendency for higher-peak-luminosity SNe IIn to occur in lower-metallicity environments \citep{2023Moriya} was also examined. The bottom panel of Figure~\ref{fig:metallicity} compares the peak-magnitude distributions of metal-rich and metal-poor SNe IIn. While a very weak tendency in the suggested direction is visually present, a statistical test gives an 88\% probability that this trend is due to random sampling.

Therefore, we find no statistically significant evidence that metallicity plays a defining role in the observed environment–luminosity correlation. This might imply that the extreme pre-SN mass loss required to produce the dense CSM characteristic of SNe IIn may be governed primarily by mechanisms that are relatively insensitive to metallicity.

\section{Comparison with direct progenitor detections}\label{sec:env_prog}

Among the analyzed SNe~IIn, we have a subsample in which their progenitors have been directly detected. The subsample comprises SNe~1997bs \citep{1999VanDyk,2000VanDyk}, 2005gl \citep{2007Gal-Yam,2009Gal-Yam}, 2009ip \citep{2010Smith,2011Foley,2013Ofek}, 2010bt \citep{2018EliasRosa}, 2010jl \citep{2011Smith,2017Fox,2024Niu}, 2015bh \citep{2016EliasRosa,2017Thone,2018Boian} and 2016jbu \citep{2018Kilpatrick,2022Brennan}. These SNe have been extensively observed and studied after their explosions; their host extinctions and metallicities at SNe sites have been well determined. We take these advantages to analysis the relationship between the progenitors and their surrounding stellar populations.

\begin{figure*}[htbp] 
    \includegraphics[width=0.31\textwidth]{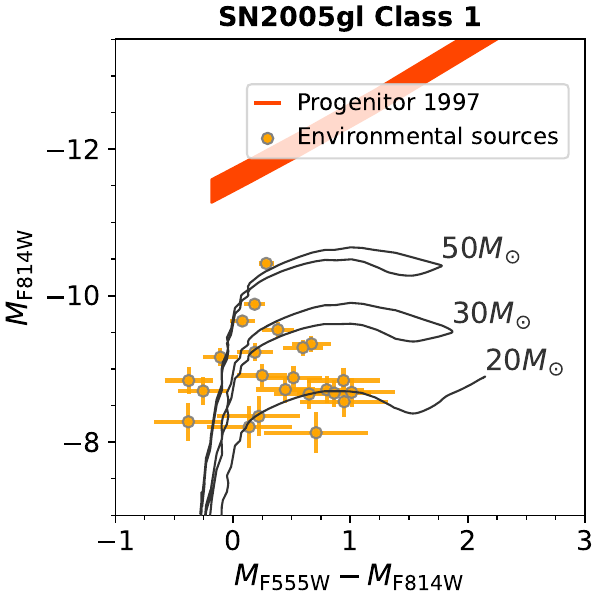}\hspace{10pt}
    \includegraphics[width=0.31\textwidth]{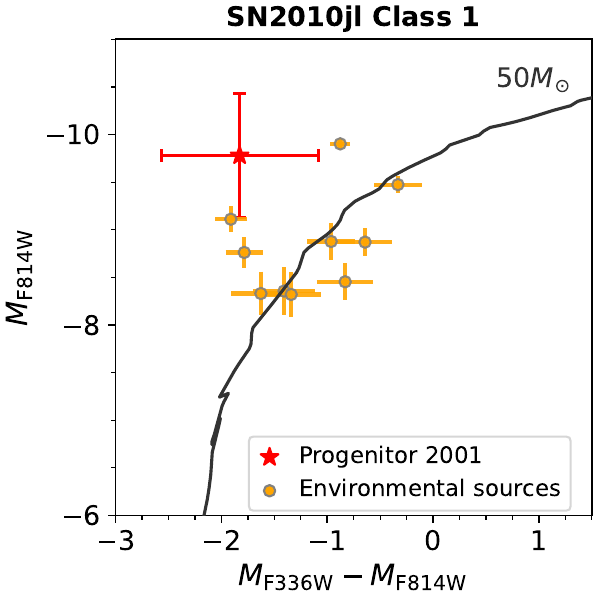}\hspace{10pt}
    \includegraphics[width=0.31\textwidth]{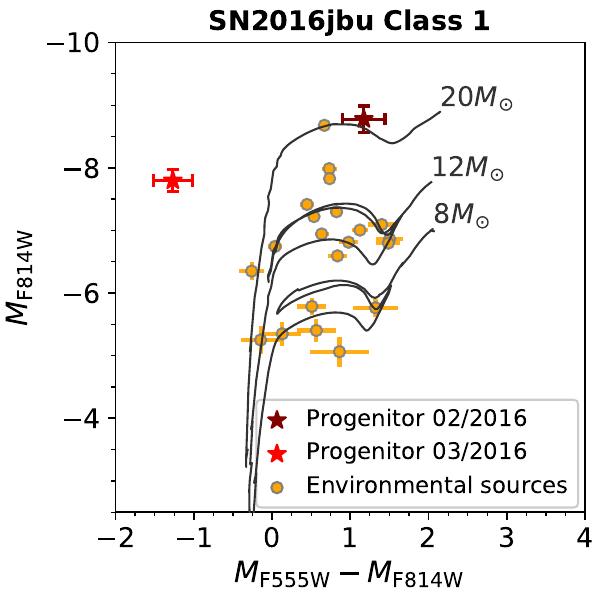}\hspace{10pt}

    \includegraphics[width=0.31\textwidth]{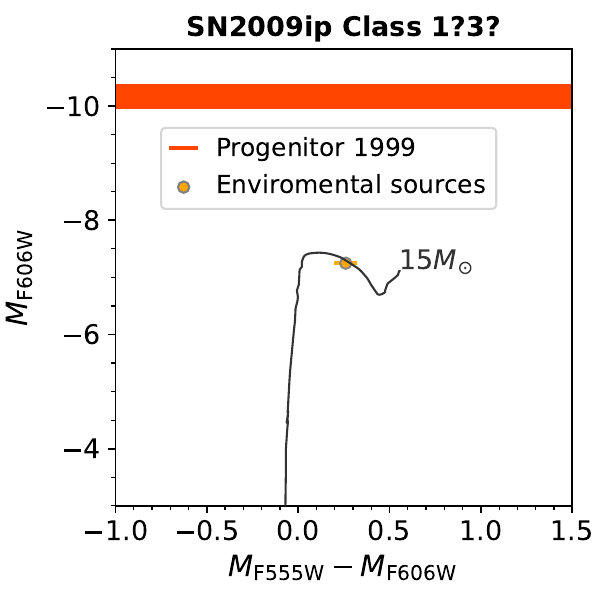}\hspace{10pt}
    
    \includegraphics[width=0.31\textwidth]{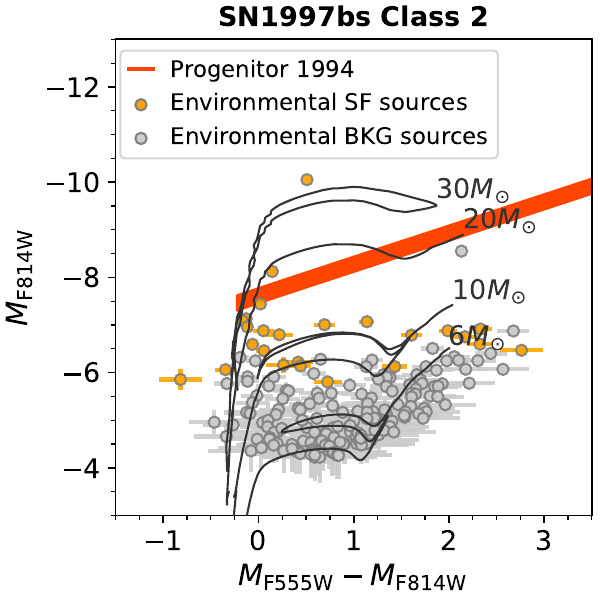}\hspace{10pt}
    \includegraphics[width=0.31\textwidth]{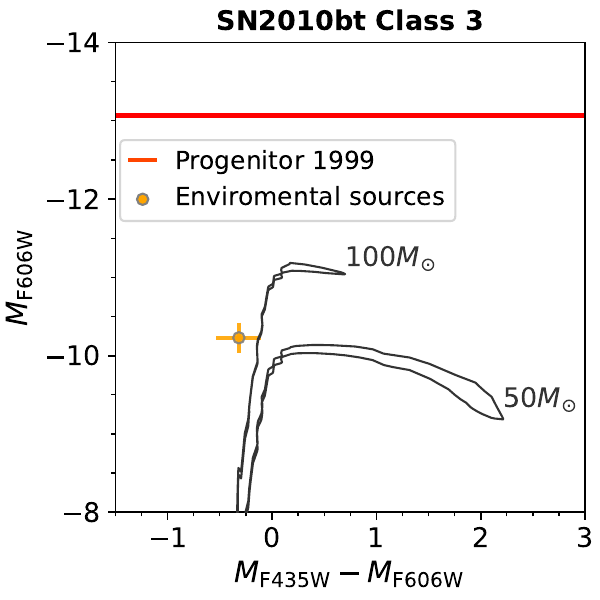}\hspace{10pt}
    \includegraphics[width=0.31\textwidth]{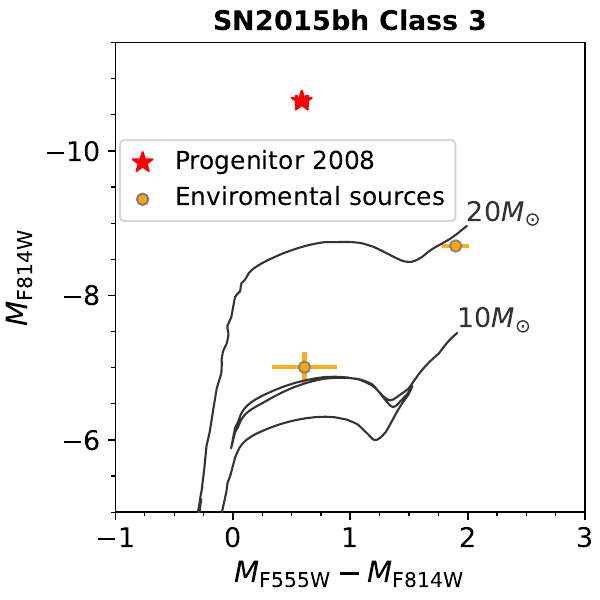}\hspace{10pt}
    \caption{Color-absolute magnitude diagrams of the resolved stellar populations surrounding the site of seven SNe~IIn with progenitor detections. 
    Each panel is labeled with the SN name and its environmental Class.
    The directly detected progenitors are marked in red, with asterisks for well-constrained detections and horizontal stripes for less certain ones.
    For Class~1, the progenitors are compared to stars in the nearest star-forming region.  
    For Class~2, we plot stars within 300~pc of the SN site. The star-forming population is shown in orange, and the background population is shown in gray (see text). 
    The Class~3 environments lack obvious star-forming regions, therefore surrounding stars within 300~pc are shown.
    The classification of SN~2009ip is uncertain (see discussion in the text). The only point source detected within 300 pc of its position has been studied in detail by \citet{2016Smith}.
    Corrections have been applied for both Galactic extinction (SFD map) and host extinction (in literature) for the progenitor and surrounding stellar.
    The widths of stripes and errorbars represent 1$\sigma$ uncertainties. 
    \textsc{parsec} single-star isochrones, with labelled initial masses, are overlaid.}
    \label{fig:prog_env}
\end{figure*}

Figure~\ref{fig:prog_env} presents the color-absolute magnitude diagram (CMD) for SNe progenitors and resolved point sources around them that are detected in both photometric bands. 
The reported progenitors of SNe~2010jl, 2015bh, and 2016jbu were detected in multiple bands, including bands that are used for environmental analysis in this work. The progenitor of SN~2010bt was detected in ACS/F330W band in 2003 and WFPC2/F606W band in 1994; due to a 9-years difference and the common occurrence of variability in possible progenitors, only the F606W-band magnitude is plotted as a horizontal line, with width representing 1$\sigma$ uncertainty. Similarly, SN~2009ip's progenitor was only observed in WFPC3/F606W band prior to the explosion and is also indicated by a horizontal line. The progenitors of SNe~1997bs (WFPC2/F606W) and 2005gl (WFPC2/F547M) were observed in bands differing from those for the environmental analysis. We interpolate progenitor magnitudes to the CMD bands using a blackbody SED with effective temperatures spanning 10$^{3.3}$–10$^{5.0}$~K and plot them in stripes.

These seven SNe~IIn cover all environment classes. SNe~2005gl, 2010jl, and 2016jbu are considered to be the Class~1 objects, and we plot resolved stars within their likely birth star-forming regions. We have reclassified SN~2009ip as Class~1 based on the discussion above; however, only one source is detected within 300~pc to the southeast of SN~2009ip. This point source has been studied in detail by \citet{2016Smith}.
\textsc{parsec} \citep{2012Bressan,2014Chen,2015Chen,2014Tang} single-star isochrones of various initial masses are overlaid for references. Clearly, their progenitors are brighter and/or bluer than the youngest stellar populations in the SNe environment, regardless of whether the local stellar ages are extremely young, $M_{\rm ini}\approx50~M_{\odot}$ for SNe~2005gl and 2010jl, or moderately young, $M_{\rm ini}\approx20~M_{\odot}$ for SN~2016jbu. 

SNe~1997bs belongs to the Class~2. Since it was outside but close to the stellar overdensity, we considered stars within 300~pc of the SN, and we supposed that stars outside the enhancements represent uniform background populations corresponding to the oldest stellar component. We identified the background (BKG) members within the enhanced regions by matching photometric counterparts (within 1$\sigma$ color and magnitude uncertainties) from the outside population. The remaining members are considered to be from the star-forming (SF) populations, representing younger stellar populations. In Figure~\ref{fig:prog_env}, the SF and BKG populations are plotted in orange and gray, respectively. 
The SF population in the vicinity exhibits $M_{\rm ini}>$10~$M_\odot$, contrasting with the lower-mass ($M_{\rm ini}<10~M_\odot$) BKG population.
We note that the progenitor is anomalous among SNe~IIn. This peculiar faintness may be physically linked to the low SN luminosity as we discussed in Section~\ref{sec:faint}. If the progenitor belonged to the BKG population, its high brightness would align with relationships observed for other Type~IIn~progenitors and their surrounding stars; alternatively, the progenitor could be fitted with single-star isochrones of $\sim$20--30~$M_\odot$.

SNe~2010bt and 2015bh were assigned Class~3, being located in a very sparse environment, lacking obvious star-forming activity. 
Stars in the vicinity of SN~2015bh are not as young as that of Class~1 and 2, with the youngest population less massive than 20~$M_\odot$. 
The resolved sources surrounding SN~2010bt are all extremely young, likely a consequence of the short exposure times that results in detections of only the brightest, most massive stars. Even with this limitation, the progenitor remained at least 1~mag brighter than all neighboring sources in the F606W band.

These results suggest that Type~IIn SNe progenitors stand out from their local stellar populations in terms of magnitude and/or color. This provides strong evidence that their progenitors are in a non-quiescent state and/or have experienced binary interactions (i.e. mergers or mass accretion) prior to explosion. In fact, mergers have been suggested to contribute significantly to the diversity of SNe, including SNe~IIn \citep{2014Justham,2019Zapartas,2024Schneider}. Spectropolarimetric observations have revealed persistent CSM asymmetry in many SNe~IIn (\citealp{2001Wang,2014Mauerhan,2018Bilinski,2024Bilinski}), which is most naturally explained by the interacting binaries. 
Meanwhile, the tension in initial mass/ages between the detected SNe progenitors and the stars in their environments has been used to identify the interacting binary histories of H-rich SNe \citep{2021Zapartas,2023Bostroem,2026niu}.
Therefore, the stellar parameters, particularly the initial mass and age, of SNe~IIn progenitors cannot be reliably estimated by standard single-star models. 
Additional, combined with the result from Section~\ref{sec:env_peak}, our number-limited subsample also suggests that SNe~IIn may originate from stellar populations with wide mass, not exclusively from very massive stellar populations, but potentially from populations with initial masses as low as $M_{\rm ini}<20$~$M_\odot$.

\section{Summary}\label{sec:summary}

In this paper, we present a study of the local environments of 31 SNe IIn within a volume-limited sample ($z < 0.02$), utilizing high-resolution archival imaging from the HST. Their environments are diverse, and we classified them into three categories: Class 1, where the SN located within active star-forming regions; Class 2, where the SN was outside but close to star-forming regions; and Class 3, where no significant star-forming region is detected around the SN. These reveal a significant diversity of SNe~IIn environments, suggesting that SNe~IIn progenitors should have multiple evolutionary pathway with a range of initial masses and lifetimes.

Our investigation reveals a clear correlation between peak absolute magnitudes of SNe~IIn and their local environments.
The ``Bright'' SNe~IIn, with $M_{\rm peak} < -19.5$ mag, are overwhelmingly found in Class~1 environments, consistent with the proposed origin from very massive progenitors. The ``Faint'' SNe~IIn in our sample, with $M_{\rm peak} > -15.5$, are preferentially associated with older environments (Classes~2 and 3).  This supports the interpretation that such events may be genuine SN explosions from the lowest-mass massive progenitors, although other channels (e.g., migrated massive stars) are not ruled out.
Meanwhile, the ``Normal" SNe~IIn, with $-19.5 < M_{\rm peak} < -15.5$ mag, are found across all three environment classes. This diversity implies that their progenitors have different initial masses and their pre-SN mass loss could be mass-insensitive and/or due to multiple physical processes.

For seven SNe~IIn with direct progenitor detections in pre-explosion images, their progenitors were compared to resolved surrounding stellar populations. These progenitors are consistently more luminous and/or bluer than the youngest population in their vicinity. This provides observational evidence that they were either in a non-quiescent state or have experienced binary merger/gainer prior to explosion. Consequently, their physical properties cannot be reliably interpreted using standard single-star evolutionary models. Again, this is against the traditional view that SNe~IIn are exclusively from very massive stars. The inferred stellar populations have a wide range of initial masses, potentially extending down to $M_{\rm ini} < 20 M_{\odot}$. 

It should be noted that the present sample, while valuable, is inherently limited by the availability of HST archival data and our current sample represents only a fraction of the known SNe~IIn population. 
This limitation currently prevents a sophisticated statistical investigation into the potential environmental dependencies of various proposed sub-types of SNe~IIn, such as the IIn-P, the 1998S-like, the 2009ip-like events. The environment-based approach established in this work provides a clear framework for future studies. 
We anticipate that a significant expansion of the sample, leveraging future high-resolution, wide-field, and multi-wavelength (particularly in the infrared) data from observatories such as the Vera C. Rubin Observatory's Legacy Survey of Space and Time (LSST), the China Space Station Survey Telescope (CSST), and the James Webb Space Telescope (JWST) will be transformative.

\bibliography{sample631}{}
\bibliographystyle{aasjournal}

\section*{acknowledgments}

This work is supported by the Strategic Priority Research Program of the Chinese Academy of Sciences, Grant No. XDB0550300, and by the China Manned Space Program with Grant No. CMS-CSST-2025-A14.
ZXN is funded by the NSFC Grant No. 12303039. 
NCS is funded by the NSFC Grants No.12303051 and No. 12261141690. JFL acknowledges support from the NSFC through Grants No. 12588202 and from the New Cornerstone Science Foundation through the New Cornerstone Investigator Program and the XPLORER PRIZE.

We are grateful to Prof. Takashi Moriya for the helpful discussions.
ZXN also acknowledges Dr. Andrea Reguitti for kindly providing images of SN~2013gc obtained from NTT, Gemini, and SOAR.

Some of the data presented in this article were obtained from the Mikulski Archive for Space Telescopes (MAST) at the Space Telescope Science Institute. The specific observations analyzed can be accessed via \dataset[doi:10.17909/xvgb-fy54]{https://doi.org/10.17909/xvgb-fy54}.

This research has made use of the Keck Observatory Archive (KOA), which is operated by the W. M. Keck Observatory and the NASA Exoplanet Science Institute (NExScI), under contract with the National Aeronautics and Space Administration.

\appendix

\setcounter{table}{0}  
\setcounter{figure}{0}
\renewcommand\thetable{\Alph{section}\arabic{table}} 
\renewcommand\thefigure{\Alph{section}\arabic{figure}} 

\section{HST log}\label{sec:hst_log}

Table~\ref{tab:HSTtab} presents the archival HST observations used in this paper.


\begin{small}
\begin{longtable}{llllllll}
\caption{HST observations of SNe\,IIn. \label{tab:HSTtab}}\\
\hline 
\hline
SN & Dataset & Instrument & Program ID  & Band & Date(UT) & Exp.Time(s)  \\
\hline
SN1961V & IC8502010 & WFC3 & 13477 & F475W & 2013-12-13 & 1143 \\ 
  & IC8502020 & WFC3 & 13477 & F814W & 2013-12-13 & 1143 \\ 
SN1978G & JC9V54010 & ACS & 13442 & F606W & 2014-05-14 & 1100 \\ 
  & JC9V54020 & ACS & 13442 & F814W & 2014-05-14 & 1100 \\ 
SN1978K & JD9G05020 & ACS & 14786 & F435W & 2016-09-27 & 1440 \\ 
  & JD9G05010 & ACS & 14786 & F606W & 2016-09-27 & 1000 \\ 
SN1994W &  IBGT19010 & WFC3 & 12229 &  F336W & 2011-01-30 & 1800 \\
SN1994Y & IEEC57020 & WFC3 & 16239 & F555W & 2021-04-07 & 710 \\ 
  & IEEC57010 & WFC3 & 16239 & F814W & 2021-04-07 & 780 \\ 
SN1996bu & JDXK07010  & ACS & 15645 & F814W & 2019-04-24 & 2304 \\
SN1997bs & ICDM12050 & WFC3 & 13364 & F438W & 2014-02-08 & 956 \\ 
  & ICDM12060 & WFC3 & 13364 & F555W & 2014-02-08 & 1134 \\ 
  & ICDM12070 & WFC3 & 13364 & F814W & 2014-02-08 & 980 \\ 
SN1997eg & IEEC06020 & WFC3 & 16239 & F555W & 2021-03-26 & 710 \\ 
  & IEEC06010 & WFC3 & 16239 & F814W & 2021-03-26 & 780 \\ 
SN1998S & ID9624020 & WFC3 & 14668 & F555W & 2016-10-03 & 710 \\ 
  & ID9624010 & WFC3 & 14668 & F625W & 2016-10-03 & 780 \\ 
SN1999el & IERG27020 & WFC3 & 16691 & F555W & 2022-02-17 & 710 \\ 
  & IERG27010 & WFC3 & 16691 & F814W & 2022-02-17 & 780 \\ 
SN2000cl & JEY361020 & ACS & 17070 & F555W & 2023-11-01 & 760 \\ 
  & JEY361010 & ACS & 17070 & F814W & 2023-11-01 & 780 \\ 
SN2003lo & J9NI01011 & ACS &  10856 &  F435W &  2006-11-08 & 1000 \\
   & J9NI03011 & ACS & 10856  &  F555W &  2006-11-17 & 1000 \\
   & J9NI04011 & ACS &  10856  &  F606W &  2006-11-17 & 1000 \\
   & J9NI07011 & ACS & 10856  &  F814W &  2006-11-19 & 1000 \\
SN2005gl & ICVY04010 & WFC3 & 14115 & F555W & 2016-10-17 & 1194 \\ 
  & ICVY04020 & WFC3 & 14115 & F814W & 2016-10-17 & 1194 \\ 
SN2005ip & IDI112010 & WFC3 & 15166 & F336W & 2018-01-11 & 780 \\ 
  & IDI112020 & WFC3 & 15166 & F814W & 2018-01-11 & 710 \\ 
SN2006gy & ICUQ27020 & WFC3 & 14149  & F625W & 2015-10-10 & 710 \\
     & ICUQ27010 & WFC3 & 14149 & F814W & 2015-10-10 & 780 \\
SN2008fq & IEBJ35010  & WFC3 & 16287 & F275W & 2021-05-24 & 1500 \\
SN2009ip & IENP04010 & WFC3 & 16671 & F438W & 2022-06-05 & 1061 \\ 
  & IEQE03020 & WFC3 & 16649 & F555W & 2021-12-14 & 1034 \\ 
  & IENP04030 & WFC3 & 16671 & F606W & 2022-06-05 & 961 \\ 
  & IENP04020 & WFC3 & 16671 & F814W & 2022-06-05 & 1061 \\ 
SN2010bt & JF5569010 & ACS & 17310 & F435W & 2024-07-23 & 674 \\ 
 & JB4U04010  & ACS & 11575 & F606W & 2010-10-09 & 40 \\
 & JB4U04020  & ACS & 11575 & F606W & 2010-10-09 & 80 \\ 
SN2010jj & IERG39020 & WFC3 & 16691 & F555W & 2022-02-04 & 710 \\
    & IERG39010 & WFC3 & 16691 & F814W & 2022-02-04 & 780 \\
SN2010jl & IEB320020 & WFC3 & 16179 & F336W & 2020-12-29 & 710 \\ 
  & IEB320010 & WFC3 & 16179 & F814W & 2020-12-29 & 780 \\ 
SN2010jp & ICKJ05010 & WFC3 & 13787 & F555W & 2015-05-14 & 4219 \\ 
  &  ICKJ05QHQ  & WFC3  & 13787 & F814W & 2016-05-14 & 1528 \\
  & IBYB10020 & WFC3 & 13029 & F625W & 2012-12-04 & 510 \\ 
  &  IBYB10010  & WFC3 & 13029 & F814W & 2012-12-04 & 680 \\
SN2011ht & ID6U01010 & WFC3 & 14614 & F438W & 2017-03-10 & 2932 \\ 
  & ID6U01030 & WFC3 & 14614 & F555W & 2017-03-10 & 1386 \\ 
  & ID6U01040 & WFC3 & 14614 & F814W & 2017-03-10 & 1386 \\ 
SN2013gc & IENP03010 & WFC3 & 16671 & F438W & 2022-03-22 & 1061 \\ 
  & IEEC52020 & WFC3 & 16239 & F555W & 2021-10-02 & 710 \\ 
  & IENP03030 & WFC3 & 16671 & F606W & 2022-03-22 & 961 \\ 
  & IENP03020 & WFC3 & 16671 & F814W & 2022-03-22 & 1061 \\ 
SN2014es &  JF550M010  & ACS & 17310 & F435W & 2025-01-31 & 674 \\
  Gaia14ahl & IDI161020 & WFC3 & 15166 & F555W & 2019-02-06 & 710 \\
   & IDI161010 & WFC3 & 15166  & F814W & 2019-02-06 & 780 \\
SN2015bf & IEBJDI010 & WFC3 & 16287 & F275W & 2021-05-12 & 1650 \\
SN2015bh & ID9607010 & WFC3 & 14668 & F438W & 2017-02-17 & 780 \\ 
  & ID9647020 & WFC3 & 14668 & F555W & 2017-01-09 & 710 \\ 
  & ID9607020 & WFC3 & 14668 & F625W & 2017-02-17 & 710 \\ 
  & ID9647010 & WFC3 & 14668 & F814W & 2017-01-09 & 780 \\ 
SN2016jbu  & IDI141020 & WFC3 & 15166 & F555W & 2019-03-21 & 710 \\ 
  & IDI141010 & WFC3 & 15166 & F814W & 2019-03-21 & 780 \\ 
SN2017hcc & IEBJAO010  &  WFC3 & 16287 & F275W & 2021-08-22 & 1650 \\
SN2019njv & IFI002010  &  WFC3 & 17770 & F300X & 2025-04-07 & 1200 \\
SN2021adxl & IFI009010 &  WFC3 & 17770 & F300X & 2025-03-17 & 1200 \\

\hline 
\end{longtable}
\end{small}

\section{Sample Environments}\label{sec:env_appendix}

Environmental analysis of other SNe~IIn used in this work. The symbols in the images have the same meanings as in Figures~\ref{fig:class1_05gl},\ref{fig:class2_00cl}, and \ref{fig:class3_15bh}. The absence of contours indicates no detection of stellar surface density enhancements. Note that although the environments of SN~2009ip, SN~2010jp, SN~2015bf, Gaia14ahl, and SN~2017hcc are sparse, their classifications were adjusted for other reasons. For details, see Section~\ref{sec:env_peak}.

\begin{figure}[H]
\centering
\includegraphics[width=0.8\textwidth]{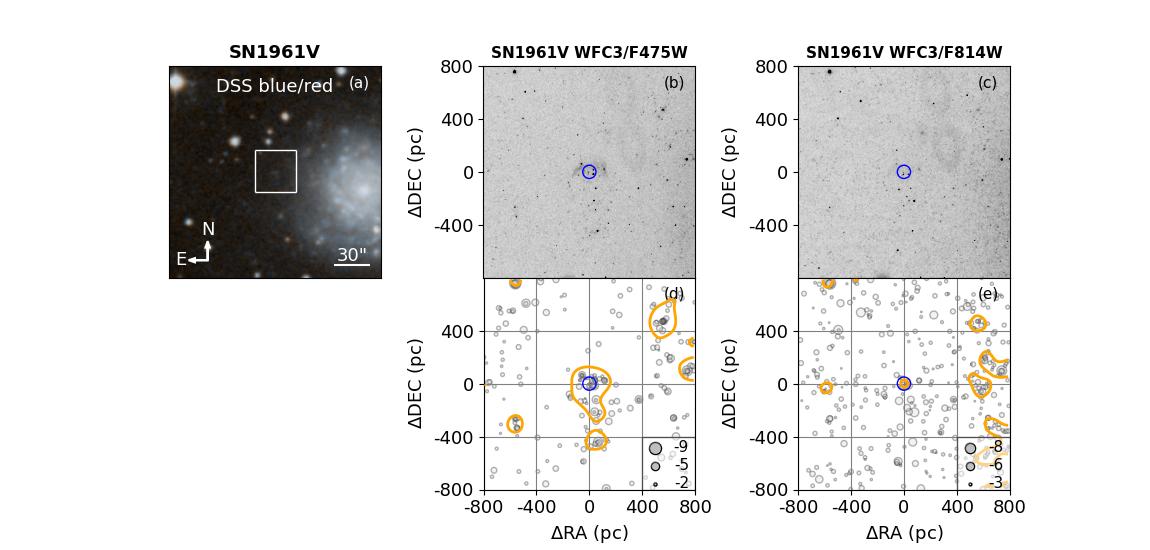}
\includegraphics[width=0.8\textwidth]{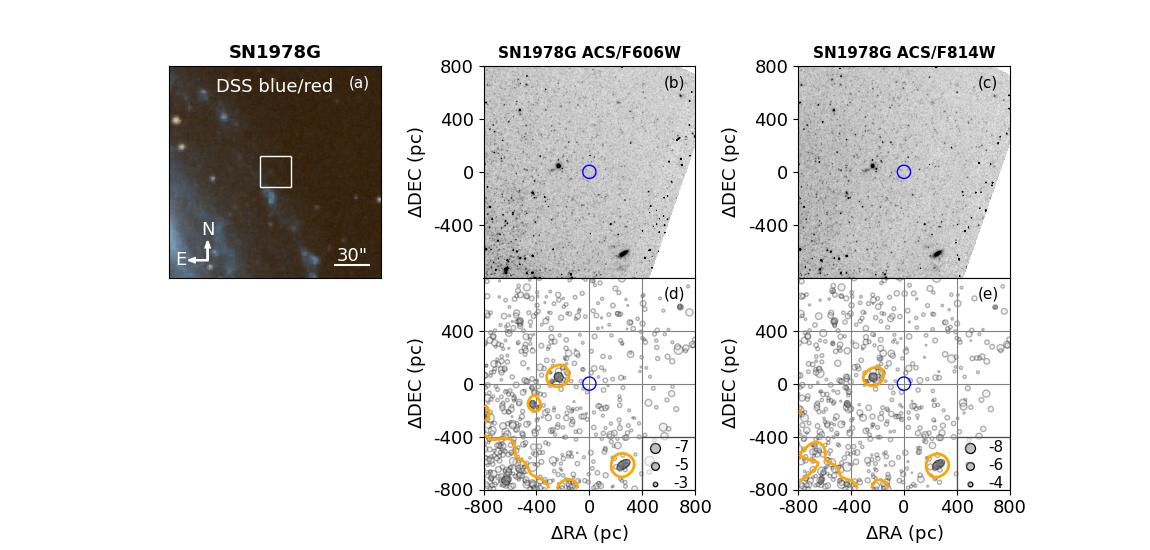}
\includegraphics[width=0.8\textwidth]{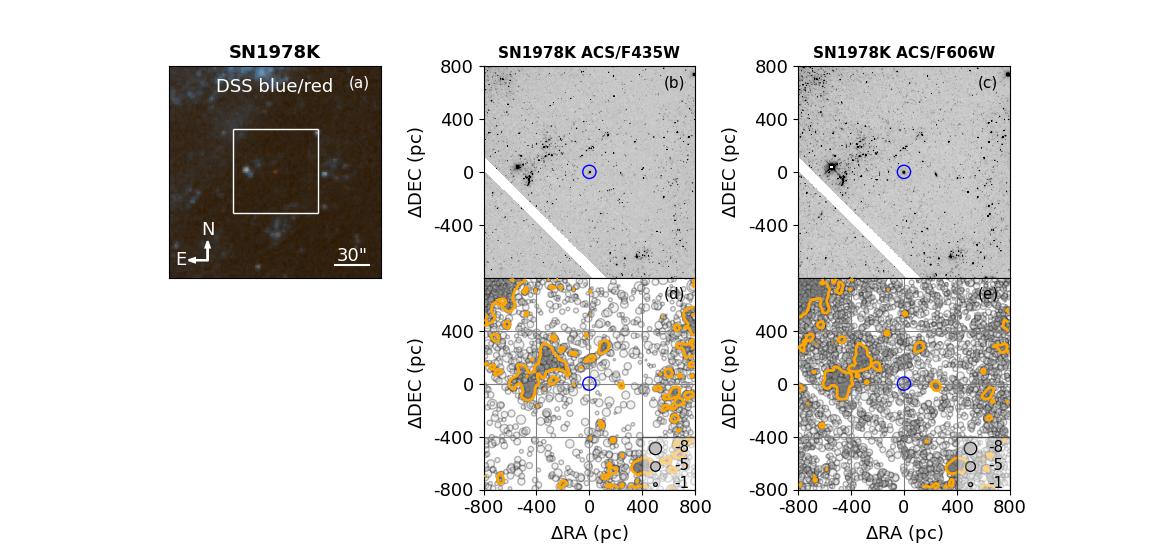}
\end{figure}
\begin{figure}[H]
\centering
\includegraphics[width=0.8\textwidth]{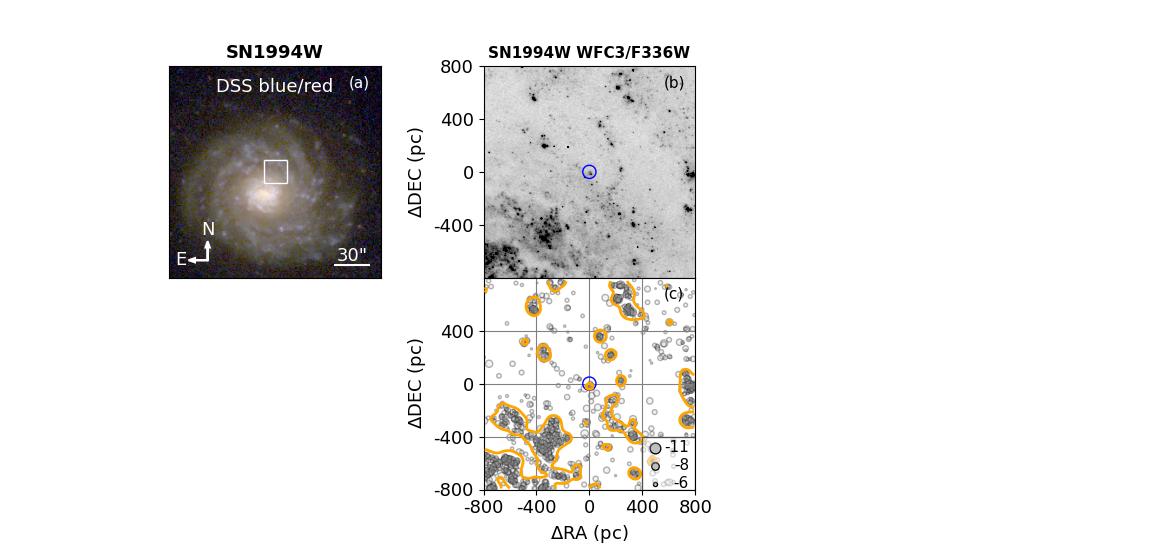}
\includegraphics[width=0.8\textwidth]{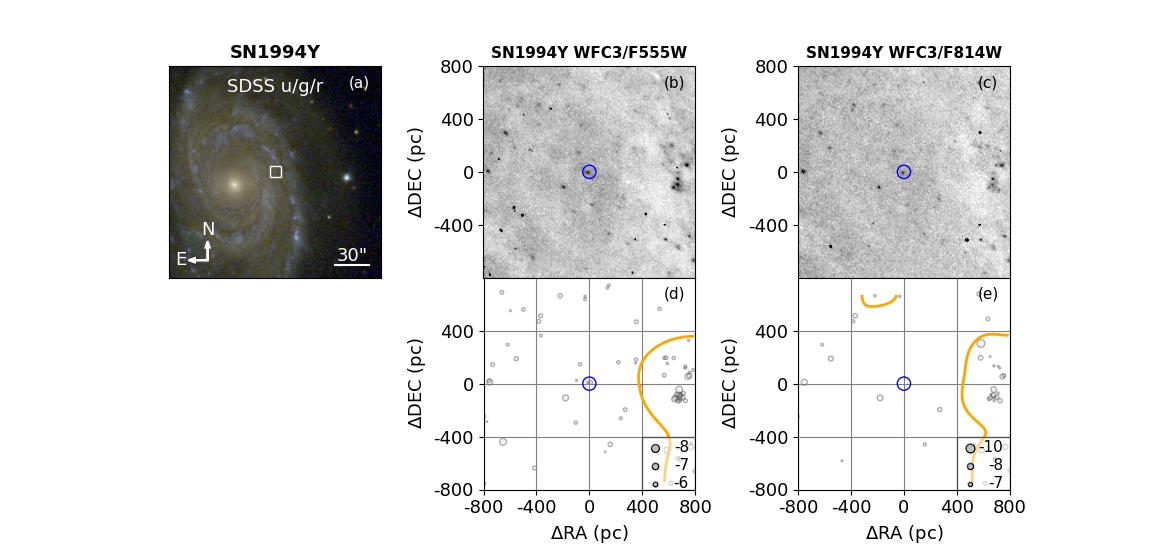}
\includegraphics[width=0.8\textwidth]{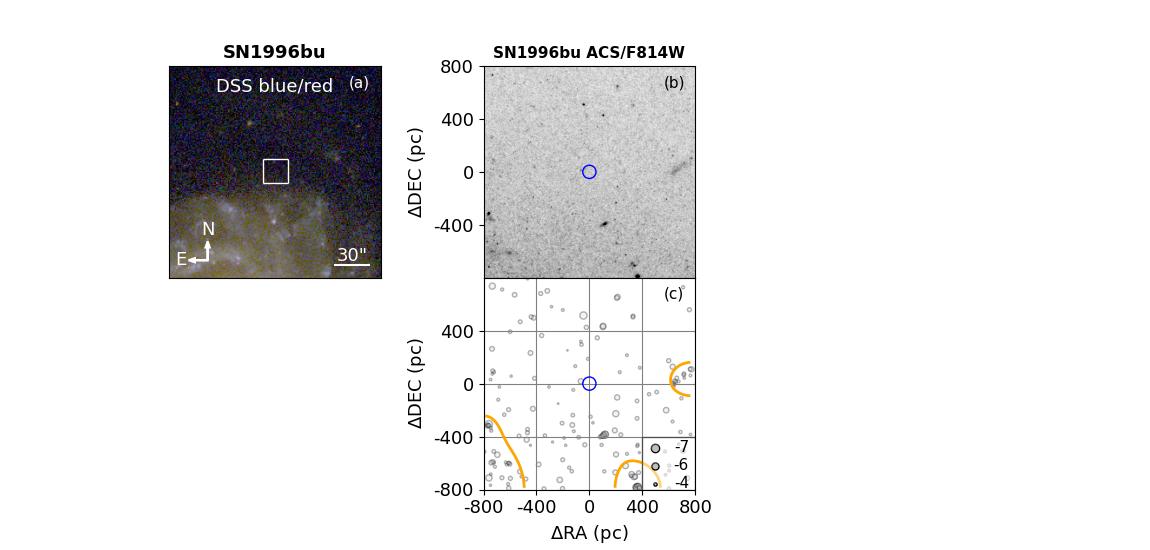}
\end{figure}
\begin{figure}[H]
\centering
\includegraphics[width=0.8\textwidth]{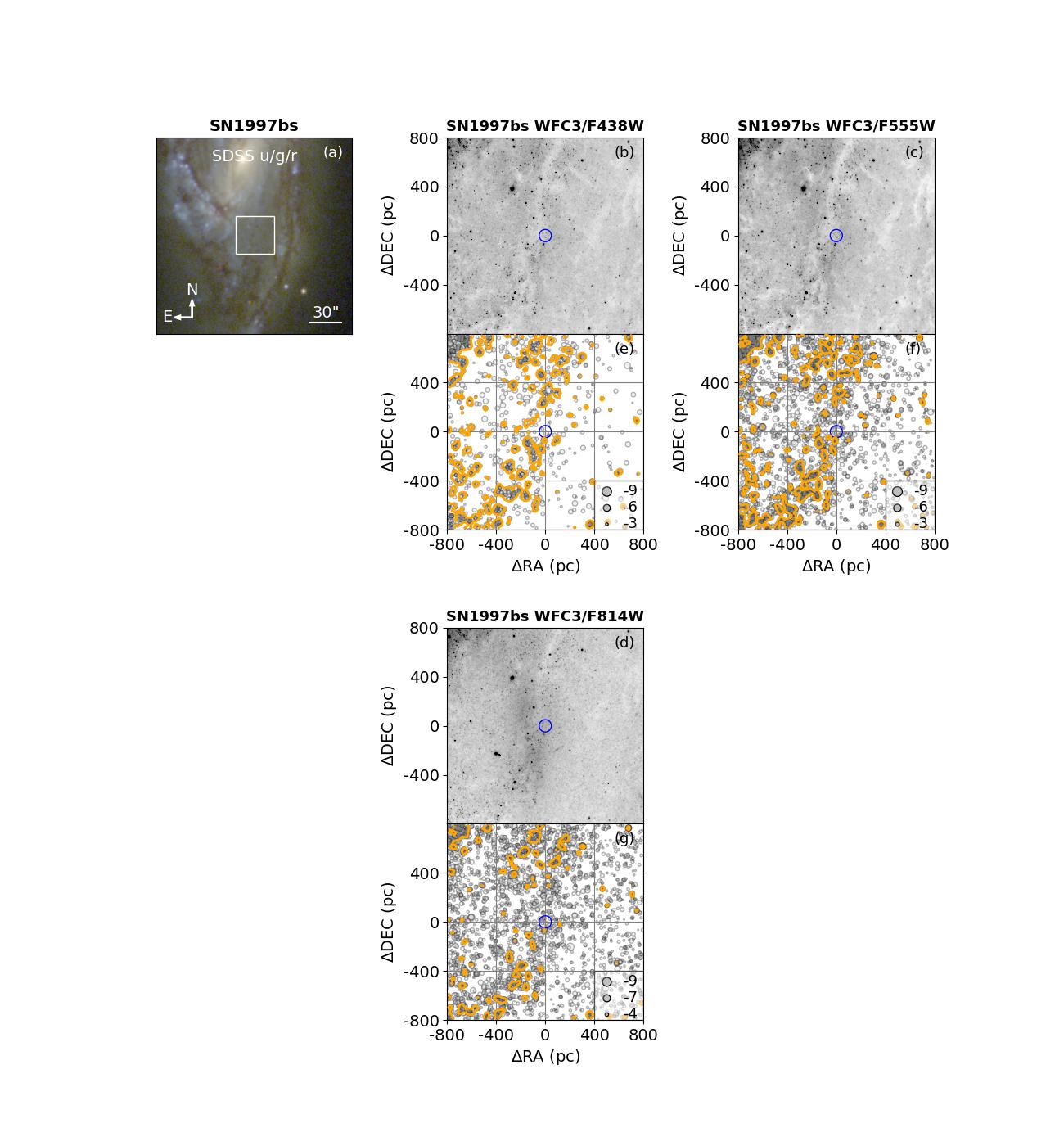}
\includegraphics[width=0.8\textwidth]{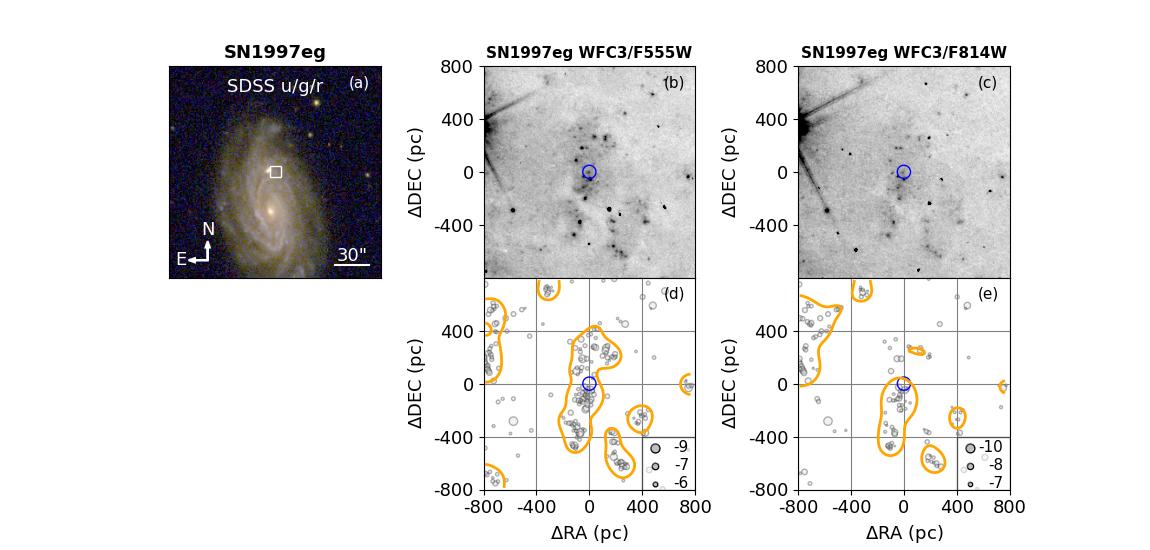}
\end{figure}
\begin{figure}[H]
\centering
\includegraphics[width=0.8\textwidth]{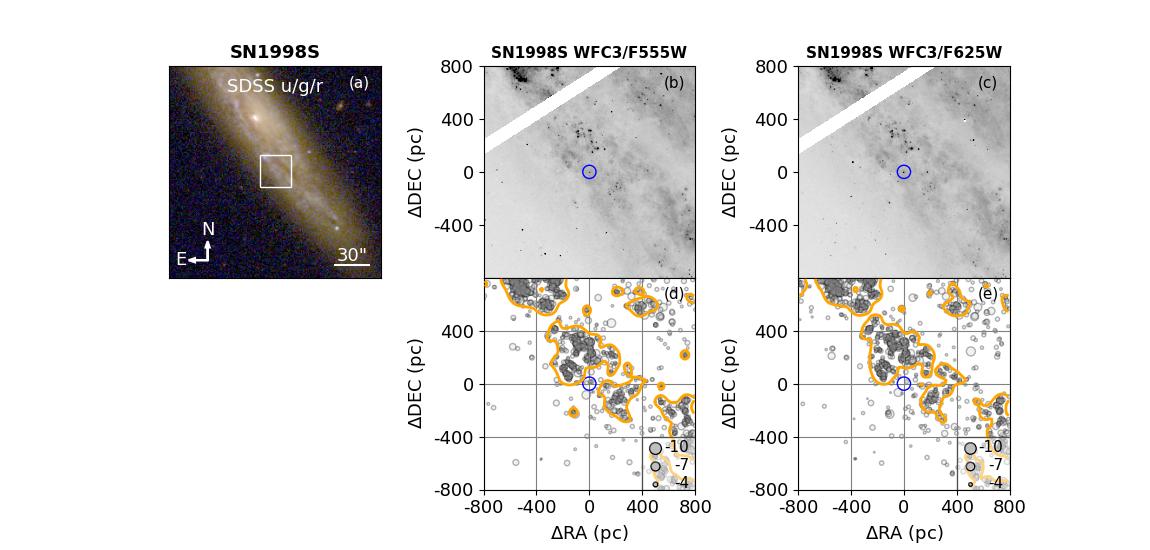}
\includegraphics[width=0.8\textwidth]{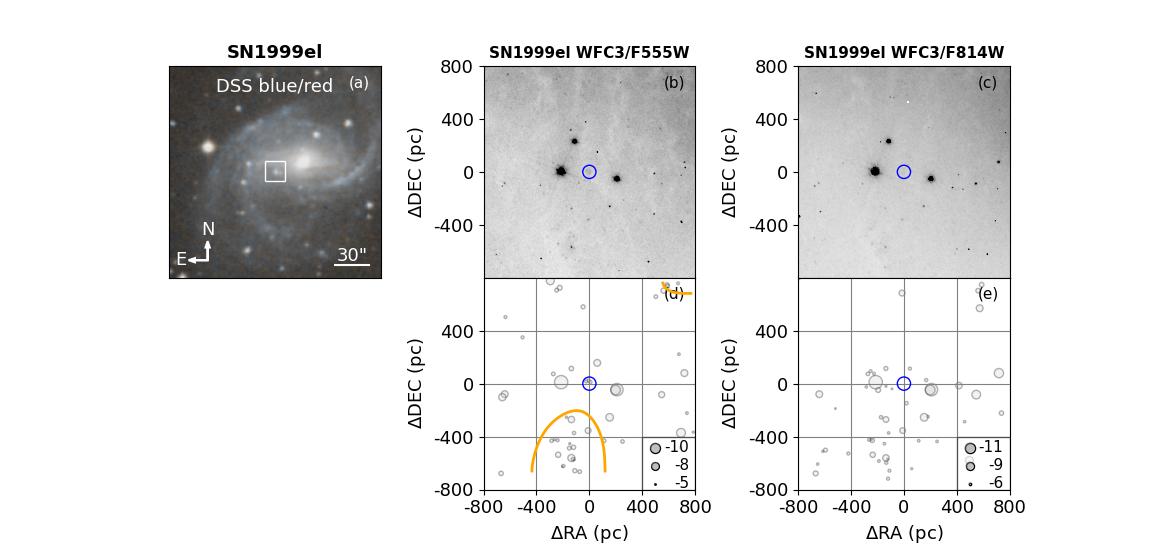}
\end{figure}
\begin{figure}[H]
\centering
\includegraphics[width=0.8\textwidth]{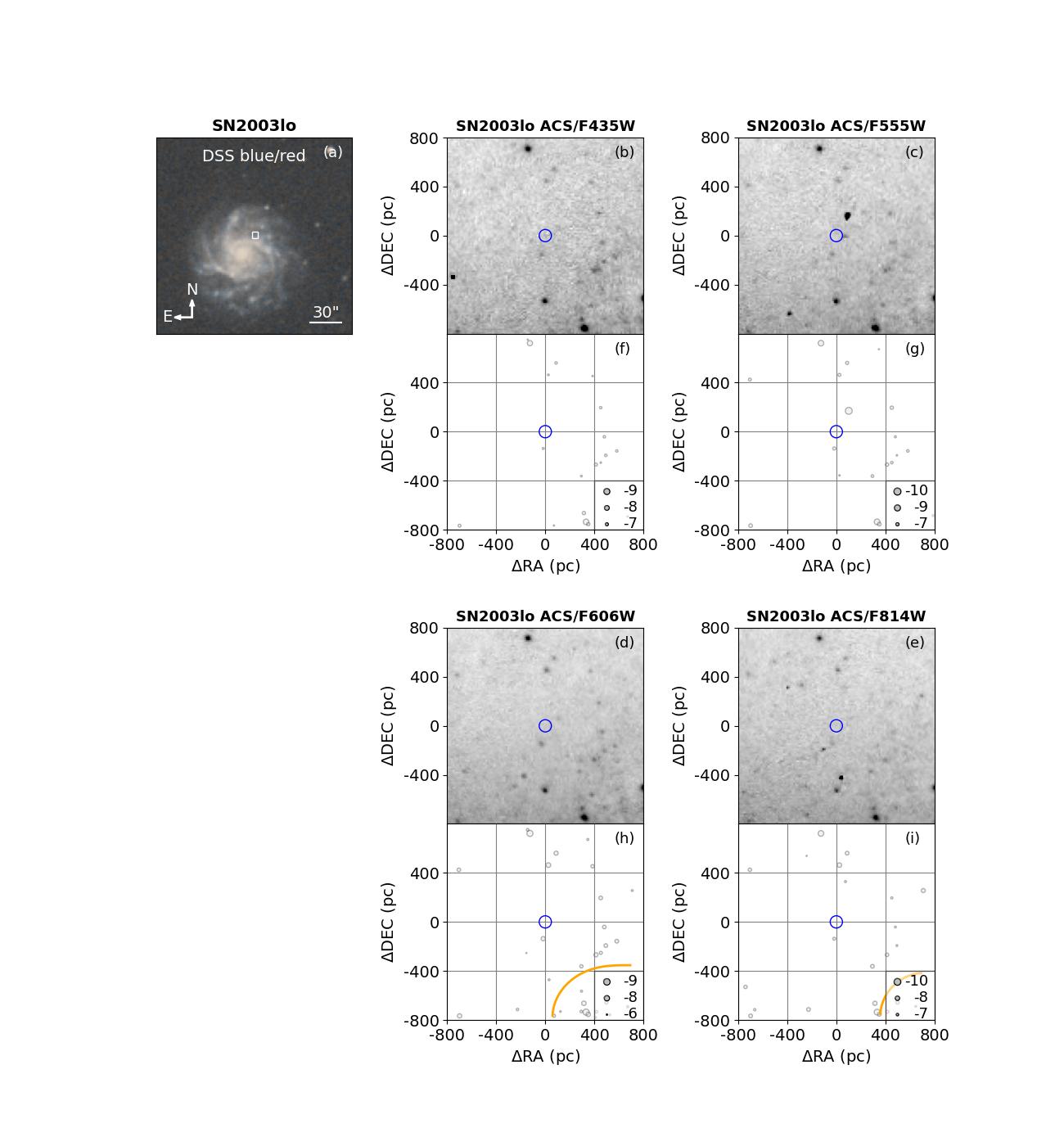}
\includegraphics[width=0.8\textwidth]{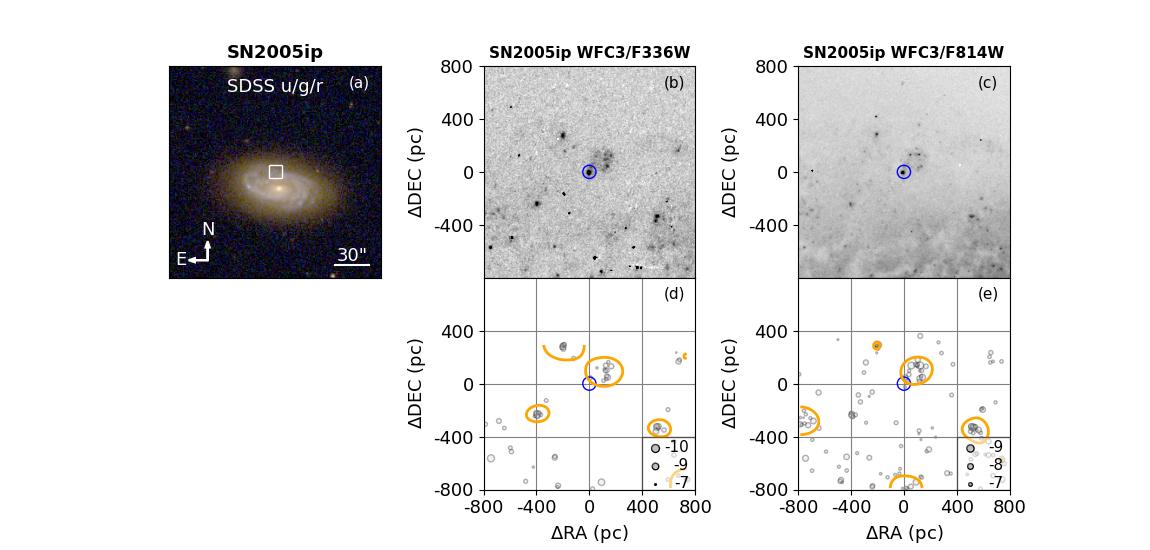}
\end{figure}
\begin{figure}[H]
\centering
\includegraphics[width=0.8\textwidth]{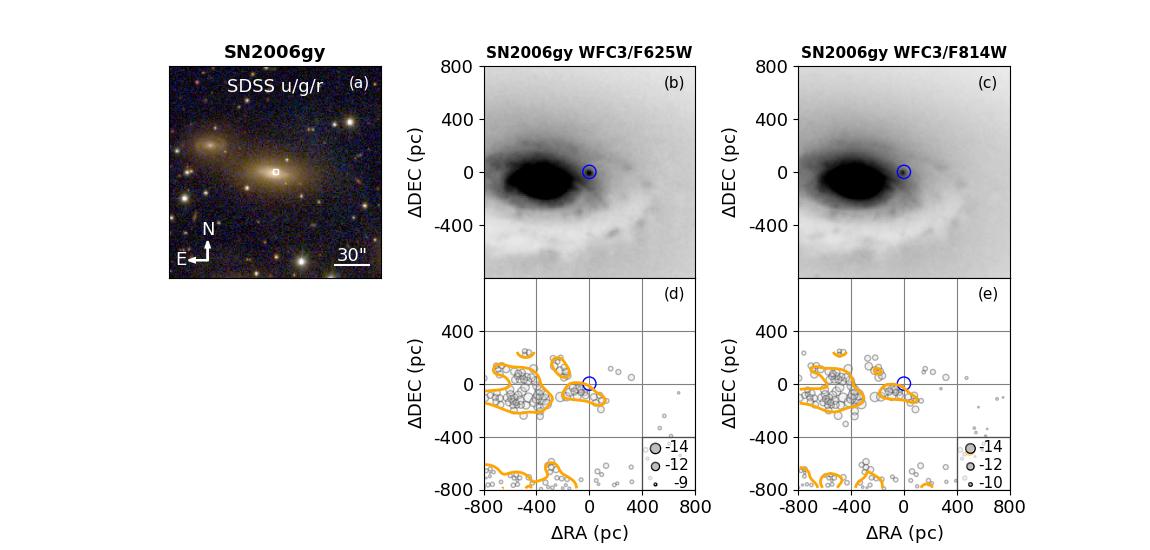}
\includegraphics[width=0.8\textwidth]{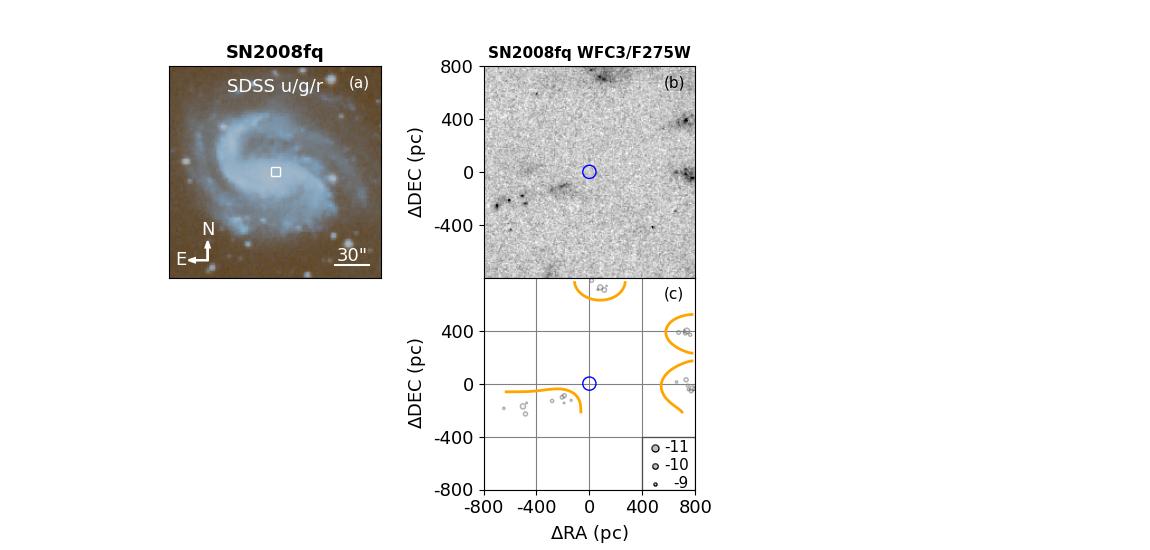}
\end{figure}
\begin{figure}[H]
\centering
\includegraphics[width=0.8\textwidth]{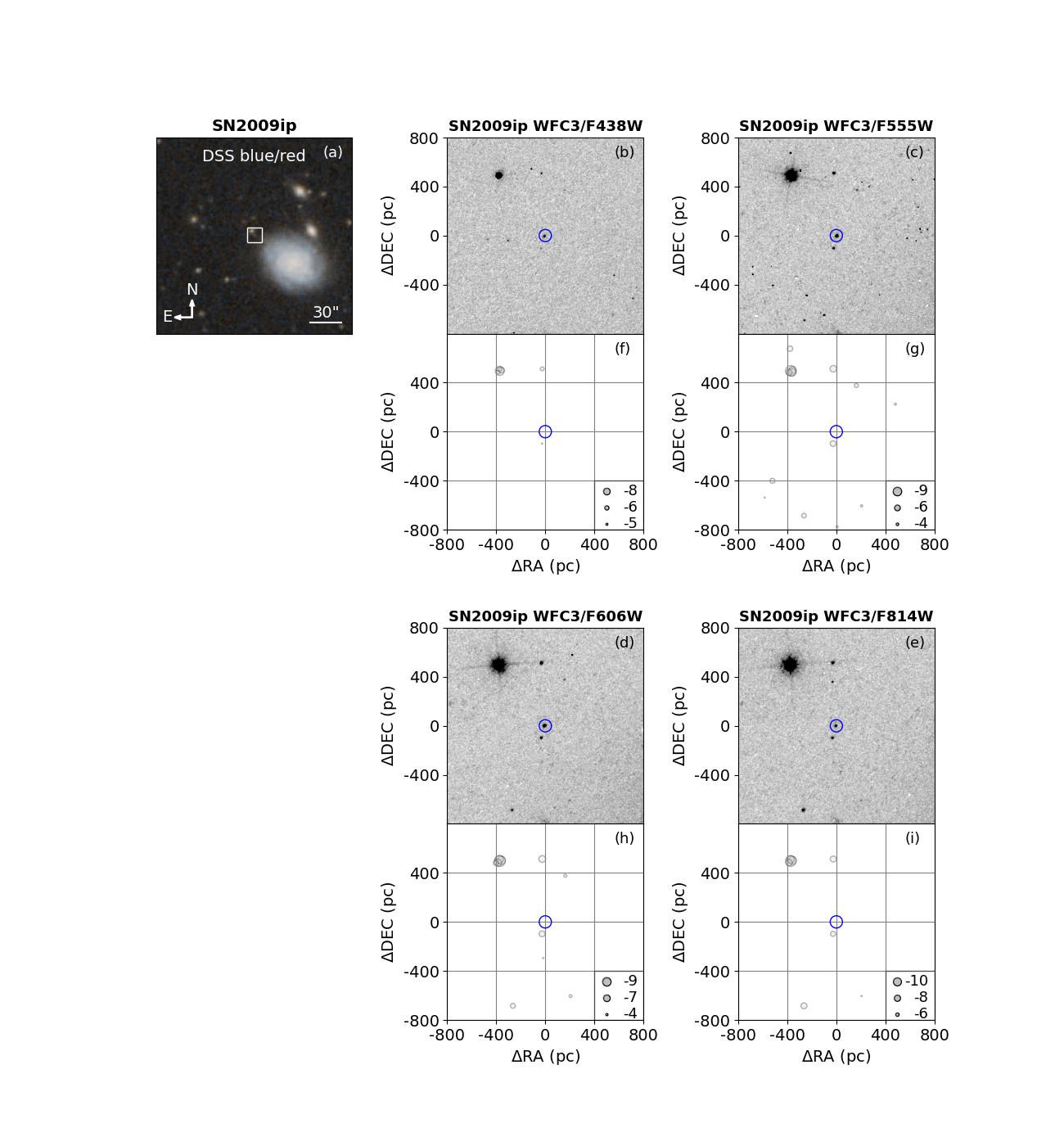}
\includegraphics[width=0.8\textwidth]{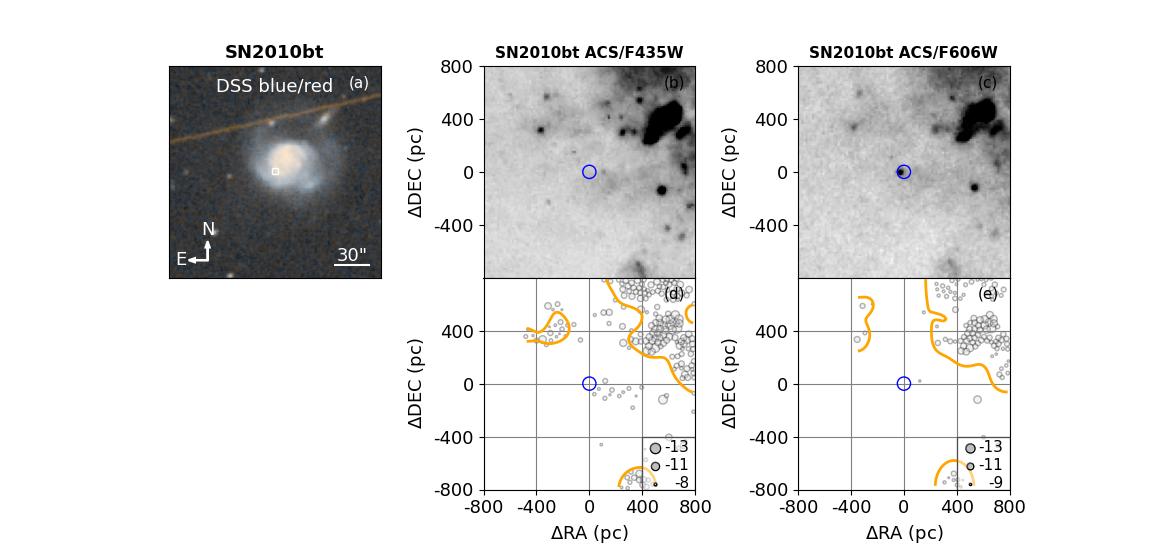}
\end{figure}
\begin{figure}[H]
\centering
\includegraphics[width=0.8\textwidth]{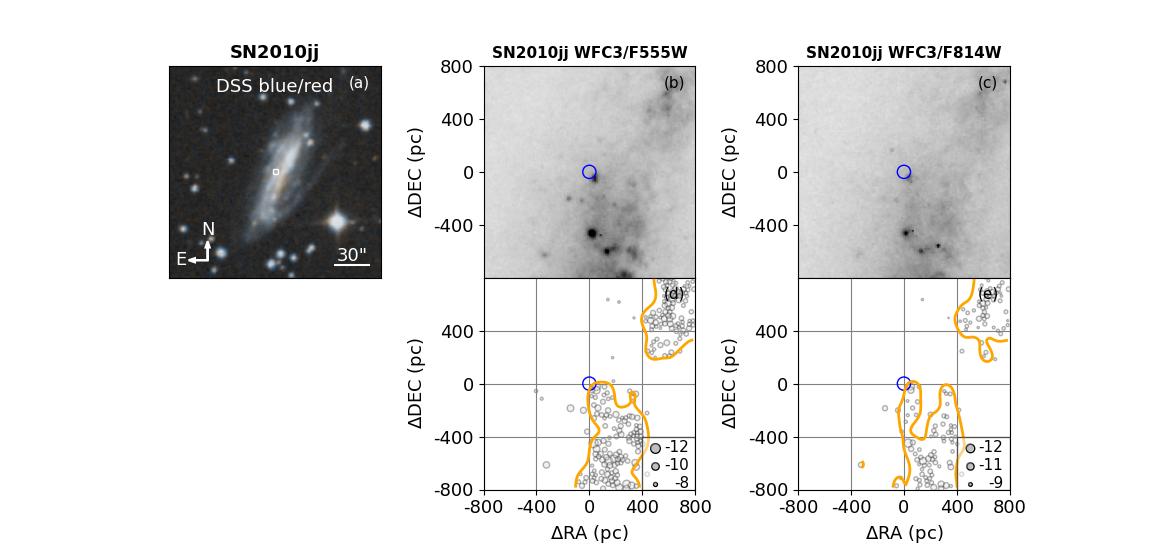}
\includegraphics[width=0.8\textwidth]{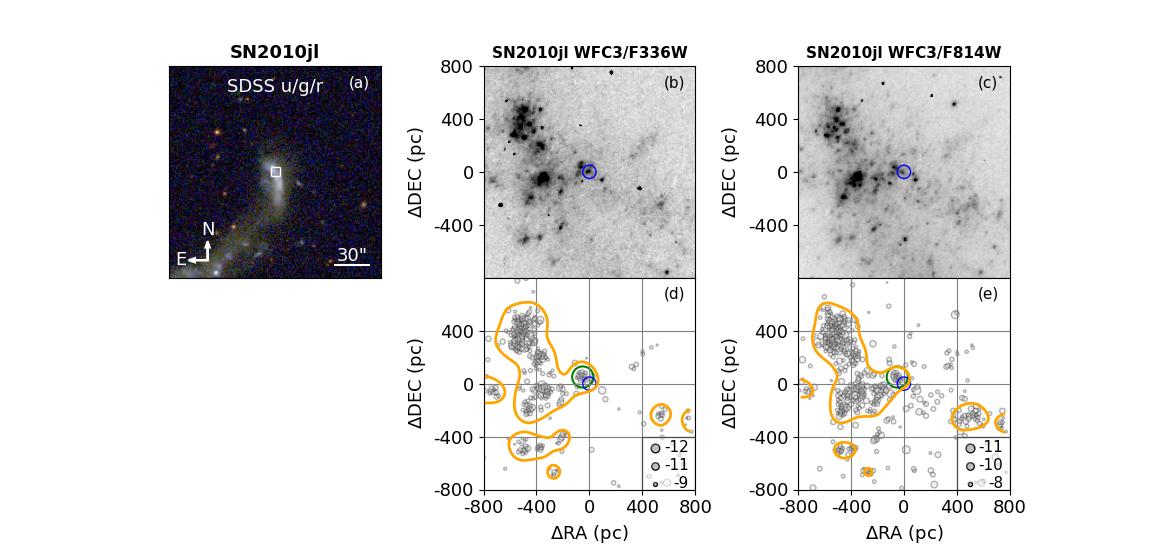}
\end{figure}
\begin{figure}[H]
\centering
\includegraphics[width=0.8\textwidth]{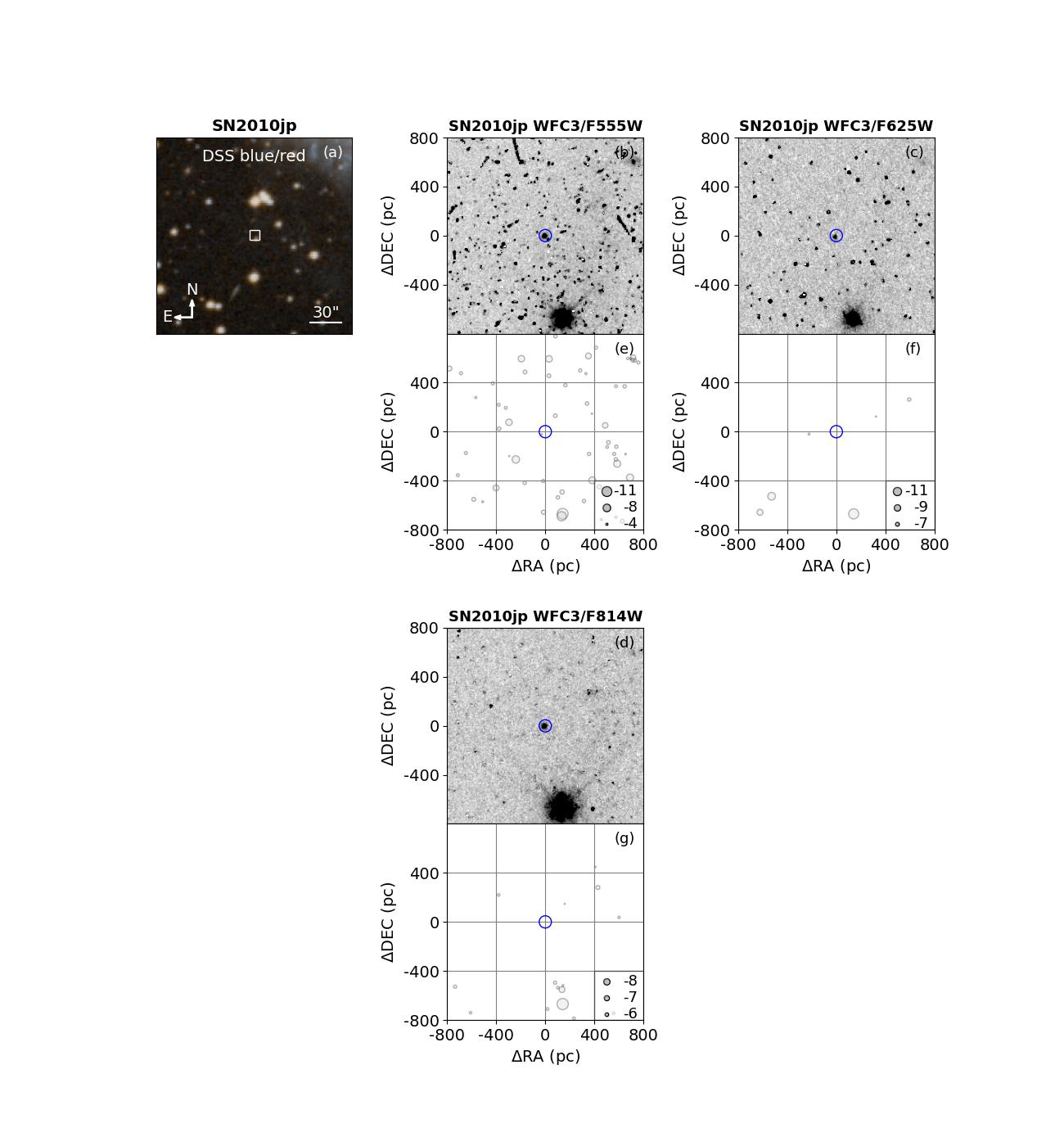}
\end{figure}
\begin{figure}[H]
\centering
\includegraphics[width=0.8\textwidth]{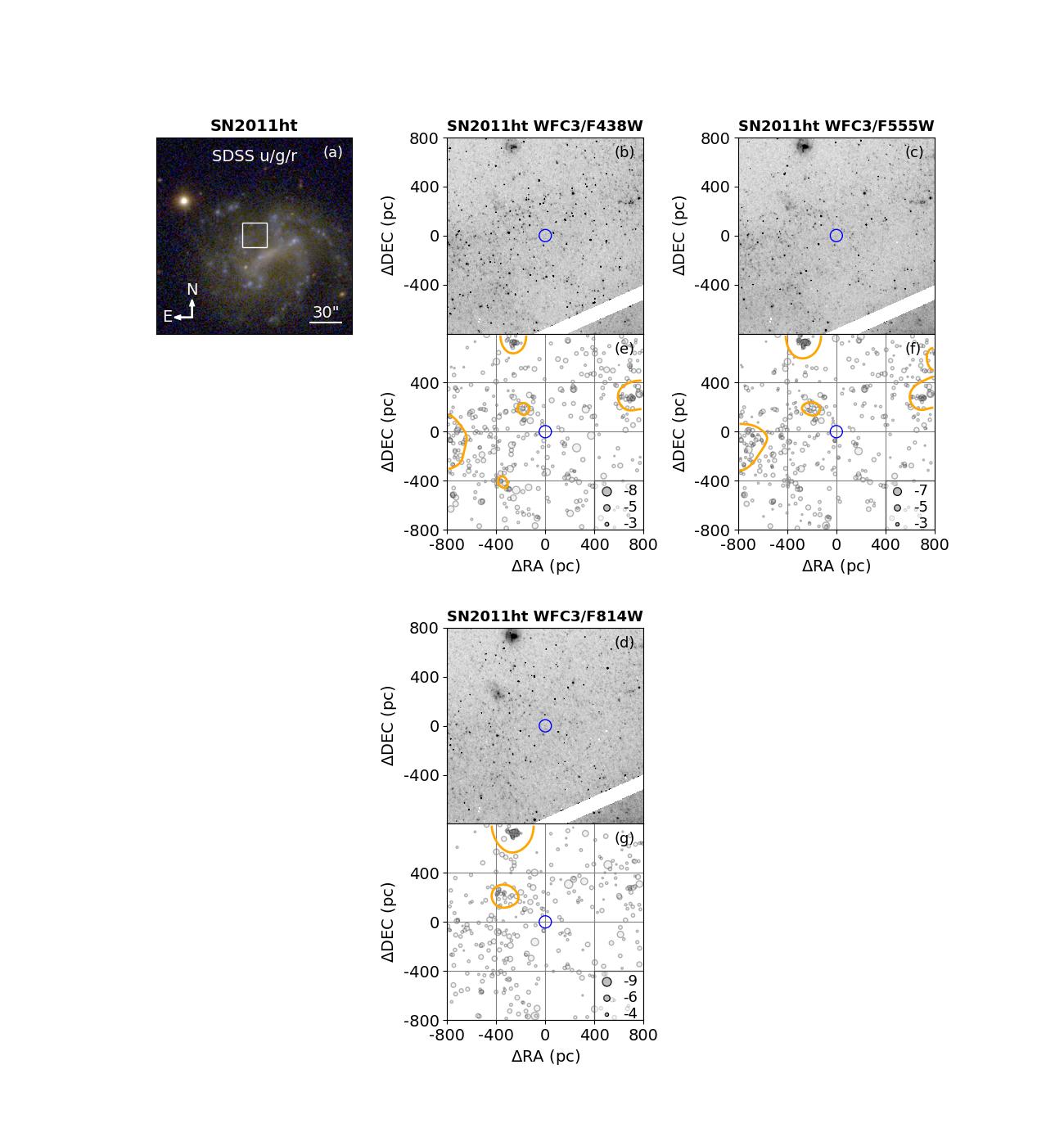}
\end{figure}
\begin{figure}[H]
\centering
\includegraphics[width=0.8\textwidth]{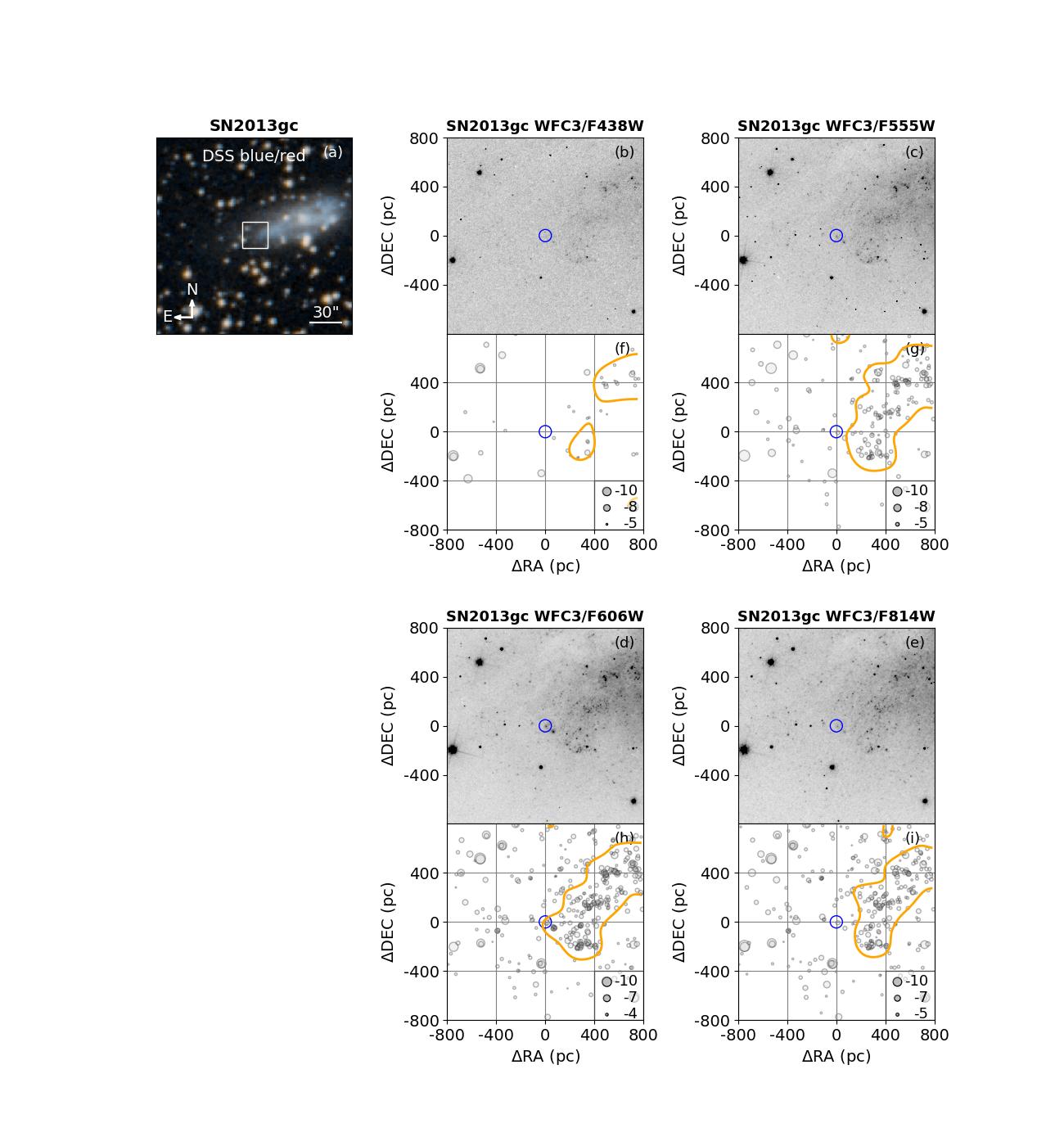}
\includegraphics[width=0.8\textwidth]{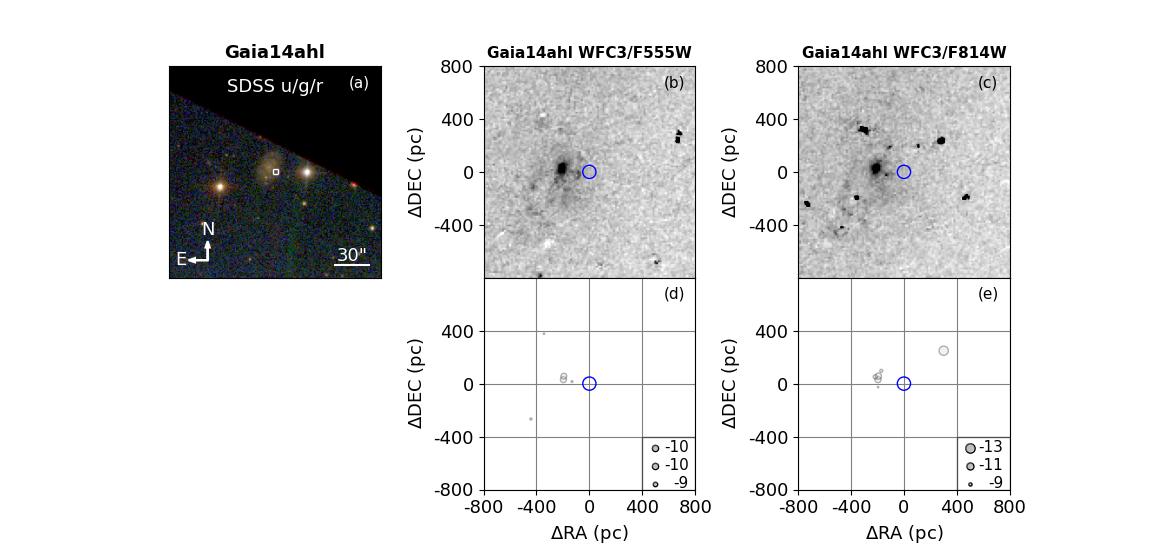}
\end{figure}
\begin{figure}[H]
\centering
\includegraphics[width=0.8\textwidth]{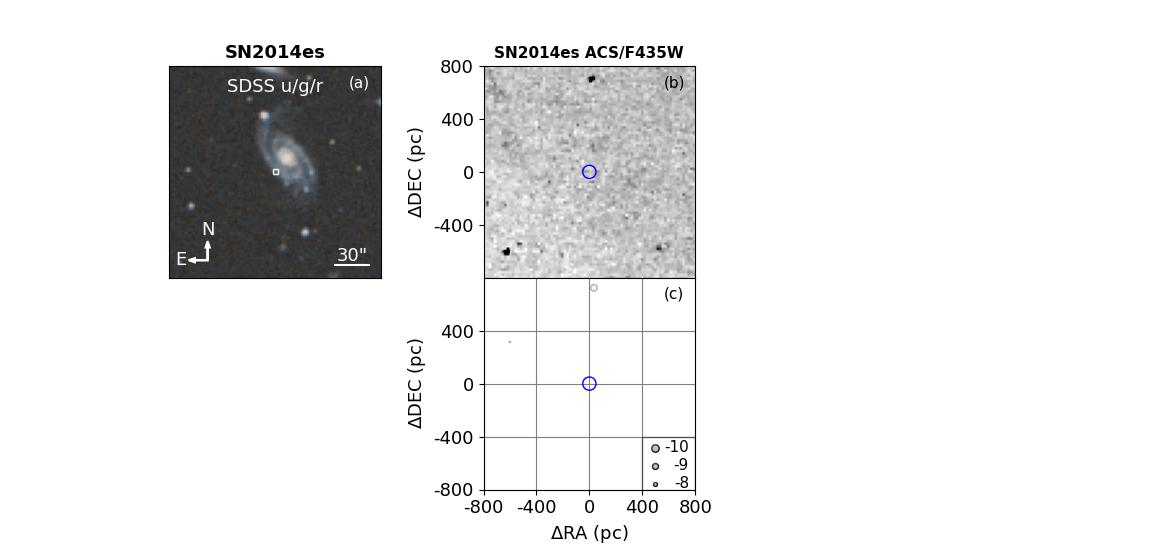}
\includegraphics[width=0.8\textwidth]{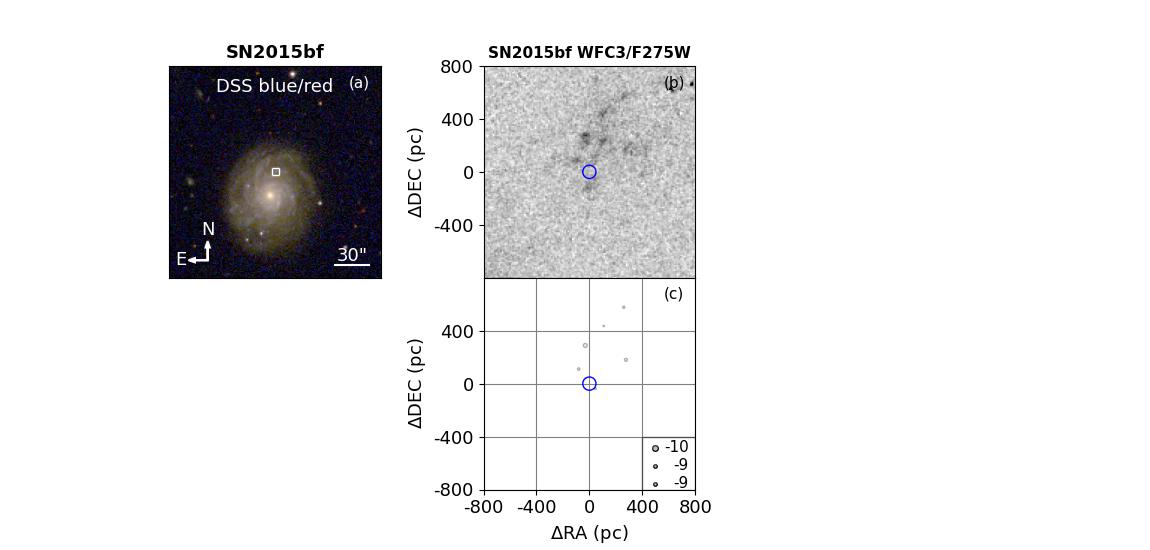}
\includegraphics[width=0.8\textwidth]{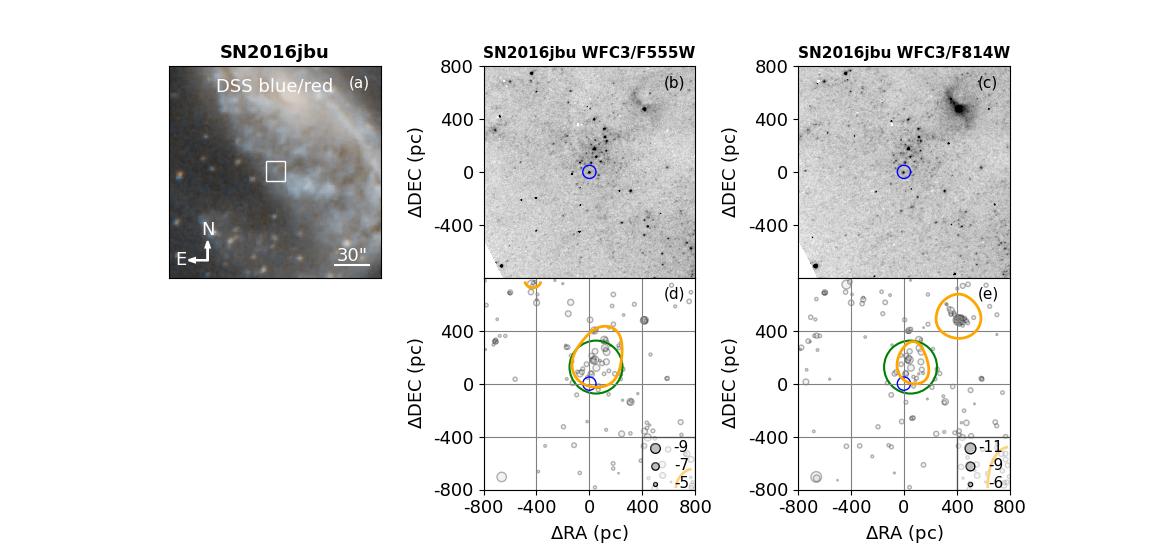}
\end{figure}
\begin{figure}[H]
\centering
\includegraphics[width=0.8\textwidth]{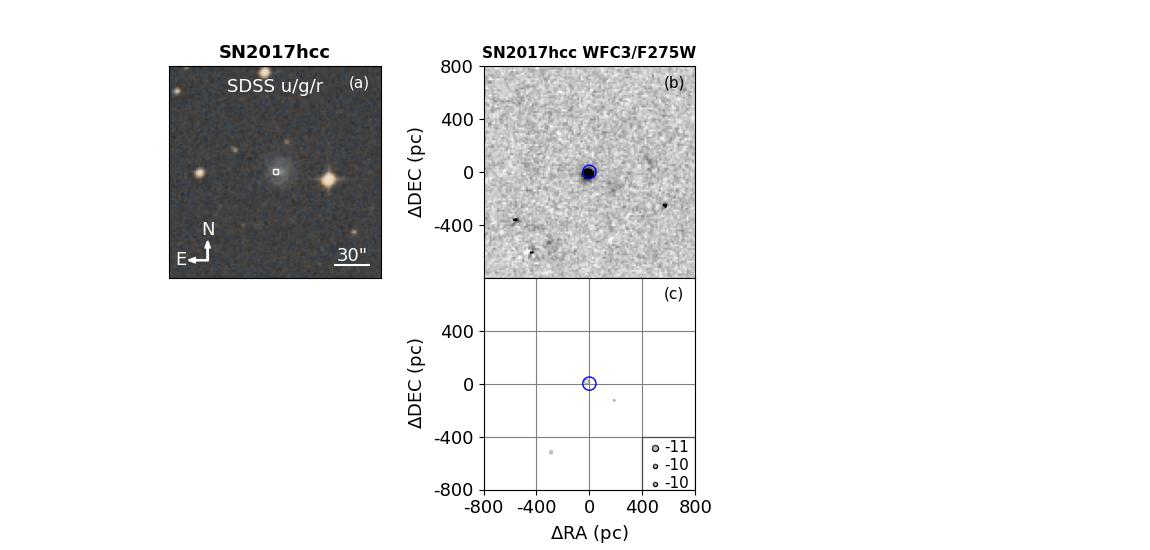}
\includegraphics[width=0.8\textwidth]{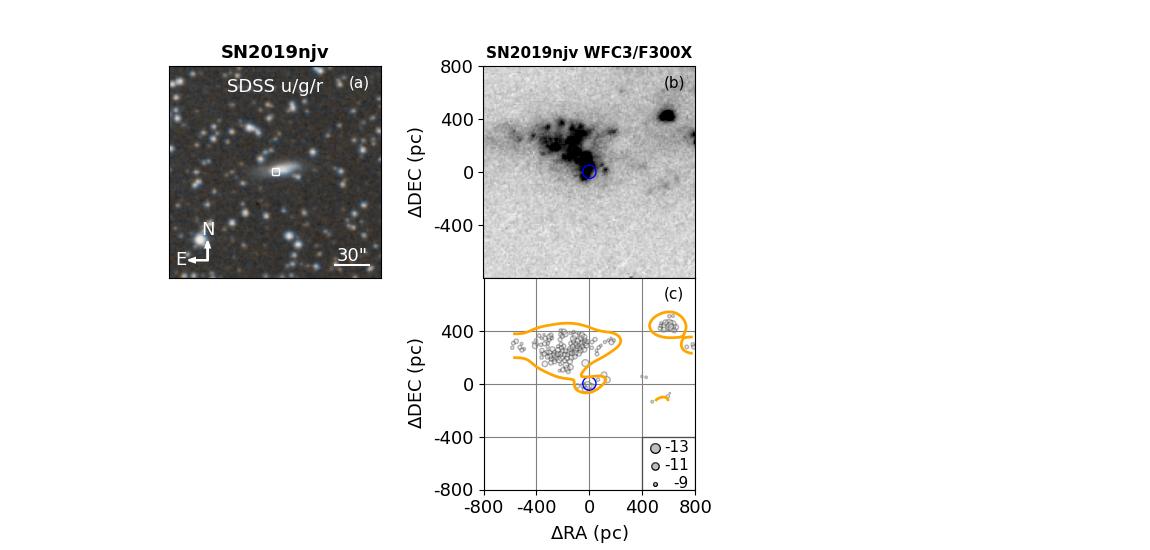}
\includegraphics[width=0.8\textwidth]{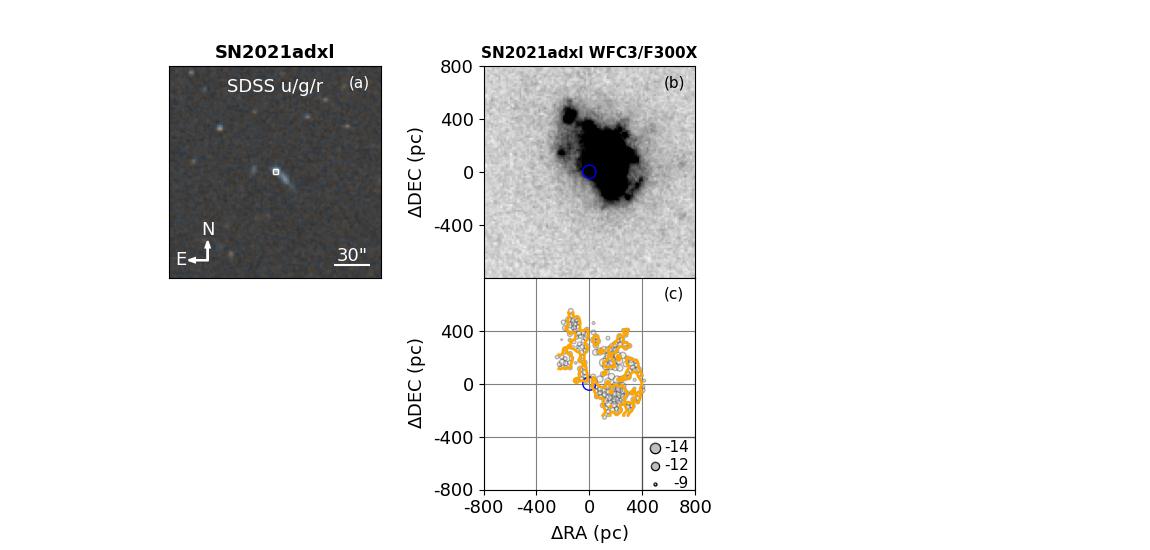}
\caption{Environmental analysis of other 28 SNe~IIn used in this work.}
\end{figure}

\section{photometric completeness}\label{sec:comp}

The HST imaging data used in this work were assembled from archival observations spanning varying filters and exposure times. We performed artificial stars tests to quantify the photometric completeness. 
For SNe with multi-band observations, their photometric completeness on the CMD plane were convoluted with the \textsc{PARSEC} stellar isochrones of given ages and host reddenings.
We assumed a total stellar mass of $10^5 M_{\odot}$ and calculated the number of recovery stars weighted by the adopted initial mass function (IMF) \citep{2002Kroupa,2013Kroupa}. 
Generally speaking, redder bands are more sensitive to older stellar population and higher extinction. Therefore, we employed the CMD from the two reddest bands available for each SN to estimate the completeness of the age and reddening. We have tested it with bluer bands and the predicted number of detected stars decreases as expected.
For SNe observed in only one band, completeness was derived solely from magnitude distributions. Most of these SNe were observed with UV bands (F300X or F275W), representing the shallowest observations in our sample.

\begin{figure}[htbp] 
    \centering %
    \includegraphics[width=0.9\linewidth]{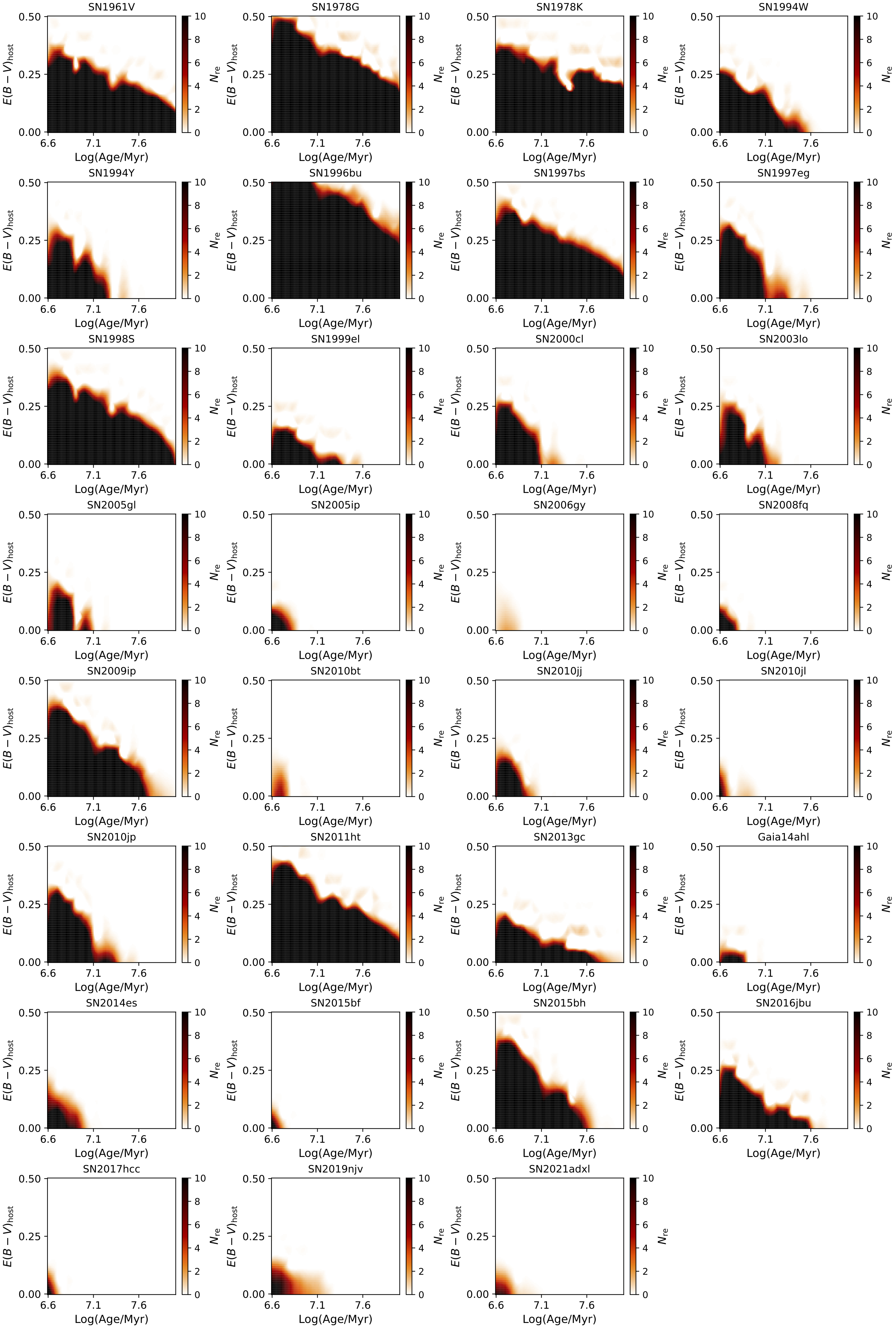}
\caption{The numbers of recovered star for a $10^5~M_{\odot}$ stellar population, shown as a function of age and reddening. }
\label{fig:comp}
\end{figure}

\end{document}